\documentclass[a4paper,11pt]{article}
\usepackage{jheppub} 
\usepackage{lineno}

\title{\boldmath Hyperfunctions in $A$-model Localization}

\author{Emil Hakan Leeb-Lundberg}
\affiliation{Department of Physics, Boğaziçi University \\
Istanbul, Turkey}
\affiliation{Niels Bohr Institute, Copenhagen University\\
Blegdamsvej 17, Copenhagen, 2100, Denmark}

\emailAdd{emilleeblundberg@gmail.com}

\abstract{We apply localization techniques to topologically $A$-twisted $\mathcal{N}=(2,2)$
supersymmetric theories of vector and chiral multiplets on $S^{2}$
and derive a novel exact formula for abelian observables, described
by a distribution integrated along the real line. The distributional
integral formula is verified by evaluating the correlator of the $A$-twisted
$\mathbb{CP}^{N-1}$ gauged linear sigma model and confirming the
standard selection rule. Finally, we use hyperfunctions to demonstrate
the equivalence between the distributional and complex contour integral
descriptions of the $\mathbb{CP}^{N-1}$ correlator, and find agreement
with the Jeffrey-Kirwan residue prescription.}

\begin{document}
\maketitle
\flushbottom

\section{Introduction \label{sec:Introduction}}
Supersymmetric quantum field theories provide a framework to investigate
mathematically tractable versions of subtle non-perturbative phenomena
occurring in nature. In this setting, supersymmetric localization
\cite{Pestun_2017} has proven crucial in evaluating quantum observables
exactly, verifying dualities, and validating other non-perturbative
techniques. Given a quantum field theory with a fermionic symmetry
$Q$, localization leverages a $Q$-exact deformation of the action
to reduce the infinite-dimensional path integral to lower-dimensional
integrals over the fixed-point locus of $Q$, facilitating the exact
evaluation of $Q$-closed observables. Different choices of $Q$-exact
deformation, or localizing term, can yield mathematically distinct,
but physically equivalent, descriptions of the same observable. In
this work, localization of topologically $A$-twisted $\mathcal{N}=(2,2)$
supersymmetric gauged linear sigma models (GLSMs) on $S^{2}$ surprisingly
results in a novel distributional integral description of observables
that, via hyperfunctions, is shown to be equivalent to the complex
contour integral description obtained through localization \cite{Benini:2015noa,Closset:2015rna}.

GLSMs \cite{Witten:1993yc} offer a concrete setting to study string
compactifications, bridge gauge theory and geometry, and perform exact
non-perturbative computations. Physically, GLSMs have yielded insights
into instantons \cite{Morrison:1994fr}, wall-crossing \cite{Gaiotto:2015aoa},
and dualities \cite{Hori:2006dk,Chen:2020iyo}. Mathematically, they
have been utilized to compute Gromov-Witten invariants \cite{Jockers:2012dk,Gomis:2012wy},
elliptic genera \cite{Benini:2013nda,Gadde:2013dda}, gamma classes
\cite{Halverson:2013qca}, and were key to the physical proof of mirror
symmetry \cite{Hori:2000kt}. Localization has been applied to $N=(2,2)$
GLSMs on the sphere $S^{2}$ \cite{Benini:2012ui,Doroud:2012xw},
the torus $T^{2}$ \cite{Benini:2013nda,Benini:2013xpa}, the hemisphere
$HS^{2}$ \cite{Hori:2013ika,Honda:2013uca} and numerous other manifolds
\cite{Benini:2015noa,Closset:2015rna,Leeb-Lundberg:2023jsj,Benini:2016hjo,Ohta:2019odi}.
Investigations of $A$-twisted $\mathcal{N}=(2,2)$ GLSMs, in particular,
have offered insights into vortices \cite{Hosomichi:2017dbc,Ohta:2019odi},
Gromov-Witten invariants \cite{Bonelli:2013mma}, Picard-Fuchs operators
\cite{Gerhardus:2018zwb}, Hori dualities \cite{Closset:2017vvl},
mirror symmetry \cite{Kim:2016jye,Ueda:2016wfa}, and the Bethe/Gauge
correspondence \cite{Nekrasov:2009uh,Nekrasov:2014xaa}.

There are at least three established approaches to localization: (i)
conventional supersymmetric localization techniques \cite{Pestun:2007rz};
(ii) supersymmetric localization techniques employing the Jeffrey--Kirwan
(JK) residue prescription (JK localization) \cite{Benini:2013nda};
and (iii) non-abelian localization techniques \cite{Witten:1992xu}.
For $A$-twisted $\mathcal{N}=(2,2)$ supersymmetric gauge theories
on $S^{2}$, conventional localization is inapplicable \cite{Benini:2015noa,Leeb-Lundberg:2023jsj};
JK localization results in observables of gauge theories with charged
matter described by an integral along a complex contour specified
by the JK residue prescription \cite{Benini:2015noa,Closset:2015rna};
and non-abelian localization results in observables of pure gauge
theories described by an integral along a real contour \cite{Leeb-Lundberg:2023jsj}.
These latter two approaches differ by a $\ensuremath{Q}$-exact term,
leading to mathematically distinct descriptions of the same observable.
Although the two descriptions are expected to be physically equivalent,
the equivalence has not been demonstrated.

This work presents a reconciliation of localization techniques for
$A$-twisted $\mathcal{N}=(2,2)$ GLSMs on $S^{2}$. We apply a stationary
phase version of supersymmetric localization \cite{Griguolo:2024ecw}
to $\mathcal{N}=(2,2)$ supersymmetric theories of vector and chiral
multiplets, topologically $A$-twisted to the $S^{2}$ using a vector-like
R-symmetry, i.e. the $A$-model on $S^{2}$. Our main result is a
novel exact formula for observables of abelian $A$-twisted $\mathcal{N}=(2,2)$
GLSMs on $S^{2}$, described by a distribution integrated along a
real contour. The formula is verified by evaluating the correlator
of the $A$-twisted $\mathbb{CP}^{N-1}$ model and confirming the
known selection rule. Then, hyperfunctions are used to relate the
distributional description of the $A$-twisted $\mathbb{CP}^{N-1}$
correlator to the established JK contour integral description \cite{Benini:2015noa,Closset:2015ohf},
providing evidence of their equivalence.

To make the discussion concrete, let us contrast the descriptions
of observables resulting from distinct approaches to localization
of $A$-twisted $\mathcal{N}=(2,2)$ GLSMs on $S^{2}$. Schematically,
these observables are described by 
\begin{equation}
\left\langle \mathcal{O}\right\rangle \propto\sum_{\mathfrak{m}}\int_{\mathcal{C}}\mathrm{d}u\;\mathcal{O}(u)\,Z_{{\rm cl}}(u,\mathfrak{m})\,Z_{{\rm 1\text{-}loop}}(u,\mathfrak{m}),\label{eq:obs}
\end{equation}
where $\mathfrak{m}$ is the quantized gauge flux, $u$ parameterizes
a bosonic scalar in the vector multiplet, $\mathcal{C}$ is the integration
contour, $\mathcal{O}$ is a gauge invariant operator insertion, $Z_{\text{cl}}$
is the classical contribution, and $Z_{\text{1-loop}}$ is the one-loop
contribution. When (\ref{eq:obs}) is the result of localization techniques
employing the JK residue prescription \cite{Benini:2015noa,Closset:2015rna},
the integration contour is the complex JK contour $\mathcal{C}=\mathcal{C}_{{\rm JK}}$,
the one-loop contribution is a meromorphic function of $u$, and integration
results in a sum over JK residues. In this work, (\ref{eq:obs}) is
derived via a stationary phase version of supersymmetric localization
\cite{Griguolo:2024ecw}, the integration contour is the real line
$\mathcal{C}=\mathbb{R}$, and the one-loop contribution is a \emph{distribution}
in $u$.

Our localization computation differs from previous examples in two key 
ways. First, our localizing term differs from those
of \cite{Benini:2015noa,Closset:2015ohf} by a $Q$-exact term that
includes a quadratic twisted chiral superpotential. Second, in evaluating
the one-loop contribution via monopole spherical harmonics on $S^{2}$,
we identify a bosonic scalar mode with purely imaginary eigenvalue.
This mode is associated to an oscillatory integral, which we evaluate
as a distribution following the treatment in \cite{Griguolo:2024ecw}.
The remaining integrals over fluctuations are evaluated as determinants.
The final one-loop contribution is a distribution multiplied by a
ratio of fluctuation determinants.

This work builds on two recent studies that revisited the original
example of non-abelian localization \cite{Witten:1992xu}. Non-abelian
localization was first applied to two-dimensional cohomological theories
of the standard multiplet, which comprises the basic, ghost, and projection
multiplets, and is related to the $A$-twisted $\mathcal{N}=(2,2)$
vector multiplet in Wess-Zumino gauge by field redefinition. The first
study, \cite{Leeb-Lundberg:2023jsj}, used field redefinition to apply
non-abelian localization techniques to $A$-twisted $\mathcal{N}=(2,2)$
theories of vector multiplets on $S^{2}$, resulting in observables
described by a function integrated along a real contour. The second
study, \cite{Griguolo:2024ecw}, introduced a stationary phase version
of localization and applied it to cohomological theories of the basic
multiplet on $S^{2}$, resulting in observables described by a distribution
integrated along a real contour. Our computation extends the one in
\cite{Leeb-Lundberg:2023jsj} by incorporating charged matter and
generalizes the techniques in \cite{Griguolo:2024ecw} to the context
of $\mathcal{N}=(2,2)$ GLSMs.

We validate our distributional formula by computing the correlator
of the $A$-twisted $\mathbb{CP}^{N-1}$ model. This is an abelian
GLSM with one vector multiplet and $N$ chiral multiplets of gauge
charge 1, which flows in the IR to a non-linear sigma model (NLSM)
with target space $\mathbb{CP}^{N-1}$. The correlator is a special
case of (\ref{eq:obs}) with insertion $\mathcal{O}(u)=u^{s}$ for
$s\in\mathbb{Z}_{\ge1}$. The correlator is expected to vanish if
the selection rule $s=N(\mathfrak{m}+1)-1$ is not satisfied, a constraint
arising from the ghost number (R-charge) anomaly described on page
34 of \cite{Morrison:1994fr}. We evaluate the correlator via the
distributional integral formula, and find agreement with established
results on page 54 of \cite{Benini:2015noa}, page 72 of \cite{Closset:2015rna},
and page 25 of \cite{Benini:2016hjo}. The correlator of the $\mathbb{CP}^{N-1}$
NLSM is discussed in, e.g., example 16.4.1 of \cite{Hori:2003ic}
and section 3 of \cite{Aspinwall:1991ce}.

A central contribution of this paper is the reconciliation of our
novel distributional integral description of observables with the
established JK contour integral description. This is achieved through
hyperfunction theory, which relates distributional integrals to complex
contour integrals. In particular, by interpreting the distributional
integrand of the $\mathbb{CP}^{N-1}$ correlator in terms of hyperfunctions,
we explicitly recover the complex contour integral description in
\cite{Benini:2015noa,Closset:2015rna,Benini:2016hjo}. Although hyperfunctions
are rarely utilized in the context of localization, they appear in
at least one six-dimensional supersymmetric localization computation
\cite{Nieri:2015dts} and in the hyperfunction formulation of Duistermaat--Heckman
localization \cite{jeffrey2017}.

The remainder of this work is structured as follows. Section 2 reviews
the necessary background, starting with $\mathcal{N}=(2,2)$ supersymmetry
on $\mathbb{R}^{2}$, followed by $A$-twisted $\mathcal{N}=(2,2)$
supersymmetry on $S^{2}$. Section 3 presents the localization computation,
including the $Q$-exact localizing term, the localization locus,
the evaluation of the chiral multiplet one-loop contribution via monopole
spherical harmonics, and the resulting distributional formula for
abelian $A$-twisted $\mathcal{N}=(2,2)$ GLSMs on $S^{2}$. Section
4 applies the results to the $\mathbb{CP}^{N-1}$ correlator, verifies
the selection rule, and demonstrates the equivalence between the distributional
and JK contour integral descriptions of observables via hyperfunctions.
Section 5 discusses implications and prospects, including non-abelian
and $\Omega$-deformed extensions.

\section{The $A$-model on $S^{2}$\label{sec:A-model}}

\paragraph{Conventions}

Our conventions follow \cite{Benini:2016qnm}. The metric on $\mathbb{R}^{2}$
is $\delta_{\mu\nu}$ ($\mu,\nu=1,2$). We use complex coordinates
$z=x^{1}+ix^{2},\;\bar{z}=x^{1}-ix^{2},$ where $\delta_{z\bar{z}}=\tfrac{1}{2}$
and $\delta_{zz}=\delta_{\bar{z}\bar{z}}=0$. Vectors obey $X_{z}=\tfrac{1}{2}(X_{1}-iX_{2}),$
$X_{\bar{z}}=\tfrac{1}{2}(X_{1}+iX_{2})$, and $X_{\mu}X^{\mu}=2(X_{z}X_{\bar{z}}+X_{\bar{z}}X_{z})$.
Anti-commuting Dirac spinors with lower indices are decomposed as
\begin{equation}
\lambda_{\alpha}=\left({\lambda_{+}\atop \lambda_{-}}\right),
\end{equation}
where $\lambda_{\pm}$ are Weyl spinors of charge $\pm1$ under $\mathrm{Spin}(2)\cong\mathrm{U}(1)_{E}$.
Spinor indices are raised and lowered as $\lambda^{\alpha}=C^{\alpha\beta}\lambda_{\beta}$
and $\lambda_{\alpha}=C_{\alpha\beta}\lambda^{\beta}$, where $C_{+-}=-C^{+-}=1$
is the antisymmetric charge-conjugation matrix, and $\lambda^{+}=\lambda_{-}$,
$\lambda^{-}=-\lambda_{+}$. Spinor indices are contracted as 
\begin{equation}
\psi\lambda=\psi^{\alpha}\lambda_{\alpha},\qquad\psi\gamma^{\mu}\lambda=\psi^{\alpha}(\gamma^{\mu})_{\alpha}{}^{\beta}\lambda_{\beta},
\end{equation}
where the euclidean gamma matrices are

\begin{equation}
\gamma^{1}=\sigma^{1},\qquad\gamma^{2}=\sigma^{2},\qquad\gamma^{3}=-i\gamma^{1}\gamma^{2}=\sigma^{3},
\end{equation}
and $\sigma^{1,2,3}$ are the Pauli matrices. We use the physics convention
with hermitian gauge fields, such that 
\begin{equation}
D_{\mu}=\partial_{\mu}-i[A_{\mu},\,\cdot\,],\quad F_{\mu\nu}=\partial_{\mu}A_{\nu}-\partial_{\nu}A_{\mu}-i[A_{\mu},A_{\nu}],\quad F_{\mu\nu}=i[D_{\mu},D_{\nu}].
\end{equation}
The math convention is recovered by $A_{\mu}\mapsto iA^{\prime}_{\mu},\;F_{\mu\nu}\mapsto iF^{\prime}_{\mu\nu}$,
for anti-hermitian $A^{\prime}_{\mu},F^{\prime}_{\mu\nu}$. A gauge
transformation generated by parameter $\vartheta$ acts as
\begin{equation}
\delta_{\vartheta}A_{\mu}=D_{\mu}\vartheta,\qquad\delta_{\vartheta}\Phi=i[\vartheta,\Phi].
\end{equation}

\subsection{$\mathcal{N}=(2,2)$ supersymmetry on $\mathbb{R}^{2}$\label{subsec:supersymmetry}}

The euclidean $\mathcal{N}=(2,2)$ vector multiplet in Wess-Zumino
gauge $\mathcal{V}$ with components $(A_{\mu},\sigma,\tilde{\sigma},\lambda,\tilde{\lambda},D_{E})$
transforms in the adjoint representation of the Lie algebra of the
gauge group $\mathfrak{g}=\text{Lie}\,G$. $\mathcal{V}$ describes
5+4 real degrees of freedom off-shell, comprising a connection $A_{\mu}$,
the bosonic scalars $\sigma$ and $\tilde{\sigma}$, the Dirac spinors
$\lambda$ and $\tilde{\lambda}$, and a real auxiliary scalar $D_{E}$.
The euclidean $\mathcal{N}=(2,2)$ chiral multiplet $\Phi$ with components
$(\phi,\psi,F)$ of $\text{U}(1)_{V}$ charge $q_{V}$ transforms
in a representation $\mathcal{R}$ of $\mathfrak{g}$, where $\text{U}(1)_{V}$
is the vector-like R-symmetry group. The anti-chiral multiplet $\tilde{\Phi}$
with components $(\tilde{\phi},\tilde{\psi},\tilde{F})$ of $\text{U}(1)_{V}$
charge $-q_{V}$ transforms in the conjugate representation $\mathcal{\tilde{R}}$
of $\mathfrak{g}$. Together, $\Phi$ and $\tilde{\Phi}$ describe
4+4 real degrees of freedom off-shell, comprising the bosonic scalars
$\phi$ and $\tilde{\phi}$, the Dirac spinors $\psi$ and $\tilde{\psi}$,
and the auxiliary scalars $F$ and $\tilde{F}$. The component fields
are complexified in euclidean signature. The generically complex scalars
$\sigma$ and $\tilde{\sigma}$ describe two real scalars $\sigma_{1}$
and $\sigma_{2}$, the generically complex spinor $\lambda$ is independent
of $\tilde{\lambda}$, and analogous statements apply to the fields
of the chiral multiplet.

The four generators of $\mathcal{N}=(2,2)$ supersymmetry are combined
as 
\begin{equation}
\delta=\frac{1}{\sqrt{2}}(\epsilon^{\alpha}Q_{\alpha}+\tilde{\epsilon}^{\alpha}\widetilde{Q}_{\alpha}),
\end{equation}
with commuting $\epsilon,\tilde{\epsilon}$, and anti-commuting $Q,\widetilde{Q},\delta$.
The supersymmetry variations of the vector multiplet are 
\begin{equation}
\begin{array}{l}
\delta A_{z}=\frac{i\epsilon_{+}\tilde{\lambda}_{+}}{\sqrt{2}}+\frac{i\tilde{\epsilon}_{+}\lambda_{+}}{\sqrt{2}},\\
\delta A_{\bar{z}}=-\frac{i\epsilon_{-}\tilde{\lambda}_{-}}{\sqrt{2}}-\frac{i\tilde{\epsilon}_{-}\lambda_{-}}{\sqrt{2}},\\
\delta\sigma=-\sqrt{2}\tilde{\epsilon}_{-}\lambda_{+}-\sqrt{2}\epsilon_{+}\tilde{\lambda}_{-},\\
\delta\tilde{\sigma}=\sqrt{2}\epsilon_{-}\tilde{\lambda}_{+}+\sqrt{2}\tilde{\epsilon}_{+}\lambda_{-},\\
\delta D_{E}=-i\sqrt{2}\epsilon_{+}D_{\bar{z}}\tilde{\lambda}_{+}+i\sqrt{2}\tilde{\epsilon}_{+}D_{\bar{z}}\lambda_{+}-i\sqrt{2}\tilde{\epsilon}_{-}D_{z}\lambda_{-}+i\sqrt{2}\epsilon_{-}D_{z}\tilde{\lambda}_{-}\\
\qquad\quad\;+\frac{i\epsilon_{-}[\sigma,\tilde{\lambda}_{+}]}{\sqrt{2}}-\frac{i\epsilon_{+}[\tilde{\sigma},\tilde{\lambda}_{-}]}{\sqrt{2}}+\frac{i\tilde{\epsilon}_{-}[\tilde{\sigma},\lambda_{+}]}{\sqrt{2}}-\frac{i\tilde{\epsilon}_{+}[\sigma,\lambda_{-}]}{\sqrt{2}},\\
\delta\lambda_{+}=\frac{\epsilon_{+}D_{E}}{\sqrt{2}}-\frac{i\epsilon_{+}F_{12}}{\sqrt{2}}+\frac{i\epsilon_{+}[\sigma,\tilde{\sigma}]}{2\sqrt{2}}+i\sqrt{2}\epsilon_{-}D_{z}\sigma,\\
\delta\lambda_{-}=\frac{\epsilon_{-}D_{E}}{\sqrt{2}}+\frac{i\epsilon_{-}F_{12}}{\sqrt{2}}-\frac{i\epsilon_{-}[\sigma,\tilde{\sigma}]}{2\sqrt{2}}+i\sqrt{2}\epsilon_{+}D_{\bar{z}}\tilde{\sigma},\\
\delta\tilde{\lambda}_{+}=-\frac{\tilde{\epsilon}_{+}D_{E}}{\sqrt{2}}-\frac{i\tilde{\epsilon}_{+}F_{12}}{\sqrt{2}}-\frac{i\tilde{\epsilon}_{+}[\sigma,\tilde{\sigma}]}{2\sqrt{2}}-i\sqrt{2}\tilde{\epsilon}_{-}D_{z}\tilde{\sigma},\\
\delta\tilde{\lambda}_{-}=-\frac{\tilde{\epsilon}_{-}D_{E}}{\sqrt{2}}+\frac{i\tilde{\epsilon}_{-}F_{12}}{\sqrt{2}}+\frac{i\tilde{\epsilon}_{-}[\sigma,\tilde{\sigma}]}{2\sqrt{2}}-i\sqrt{2}\tilde{\epsilon}_{+}D_{\bar{z}}\sigma,\\
\delta F_{12}=-\sqrt{2}\epsilon_{+}D_{\bar{z}}\tilde{\lambda}_{+}-\sqrt{2}\tilde{\epsilon}_{+}D_{\bar{z}}\lambda_{+}-\sqrt{2}\tilde{\epsilon}_{-}D_{z}\lambda_{-}-\sqrt{2}\epsilon_{-}D_{z}\tilde{\lambda}_{-}.
\end{array}\label{eq:susyR2_vpletVariation1}
\end{equation}
The supersymmetry variations of the chiral multiplet are 
\begin{equation}
\begin{array}{l}
\delta\phi=\epsilon_{+}\psi_{-}-\epsilon_{-}\psi_{+},\\
\delta\tilde{\phi}=\tilde{\epsilon}_{-}\tilde{\psi}_{+}-\tilde{\epsilon}_{+}\tilde{\psi}_{-},\\
\delta\psi_{+}=2i\tilde{\epsilon}_{-}D_{z}\phi+i\sigma\tilde{\epsilon}_{+}\phi+i\epsilon_{+}F,\\
\delta\psi_{-}=2i\tilde{\epsilon}_{+}D_{\bar{z}}\phi+i\tilde{\sigma}\tilde{\epsilon}_{-}\phi+i\epsilon_{-}F,\\
\delta\tilde{\psi}_{+}=-2i\epsilon_{-}D_{z}\tilde{\phi}-i\tilde{\sigma}\epsilon_{+}\tilde{\phi}+i\tilde{\epsilon}_{+}\tilde{F},\\
\delta\tilde{\psi}_{-}=-2i\epsilon_{+}D_{\bar{z}}\tilde{\phi}-i\sigma\epsilon_{-}\tilde{\phi}+i\tilde{\epsilon}_{-}\tilde{F},\\
\delta F=2\tilde{\epsilon}_{+}D_{\bar{z}}\psi_{+}-2\tilde{\epsilon}_{-}D_{z}\psi_{-}-\tilde{\epsilon}_{+}\sigma\psi_{-}+\tilde{\epsilon}_{-}\tilde{\sigma}\psi_{+}+\sqrt{2}\tilde{\epsilon}_{+}\tilde{\lambda}_{-}\phi-\sqrt{2}\tilde{\epsilon}_{-}\tilde{\lambda}_{+}\phi,\\
\delta\tilde{F}=2\epsilon_{+}D_{\bar{z}}\tilde{\psi}_{+}-2\epsilon_{-}D_{z}\tilde{\psi}_{-}+\epsilon_{-}\sigma\tilde{\psi}_{+}-\epsilon_{+}\tilde{\sigma}\tilde{\psi}_{-}+\sqrt{2}\tilde{\phi}\epsilon_{+}\lambda_{-}-\sqrt{2}\tilde{\phi}\epsilon_{-}\lambda_{+}.
\end{array}\label{eq:susyR2_chpletVariation1}
\end{equation}
The supersymmetry algebra is 
\begin{equation}
\{Q_{\alpha},\widetilde{Q}_{\beta}\}=(2\gamma^{\mu}P_{\mu})_{\alpha\beta},\quad\left\{ \delta_{\epsilon},\delta_{\tilde{\epsilon}}\right\} =i\mathcal{L}_{\epsilon\gamma^{\mu}\tilde{\epsilon}},\label{eq:susyR2_algebra}
\end{equation}
where $P_{\mu}=-i\partial_{\mu}$, $\mathcal{L}_{\epsilon\gamma^{\mu}\tilde{\epsilon}}$
is the Lie derivative along $\epsilon\gamma^{\mu}\tilde{\epsilon}$,
and the central charges have been set to zero.

Under $\text{U}(1)_{V}$, the supercharges ($Q_{+}$, $Q_{-}$, $\tilde{Q}_{+}$, $\tilde{Q}_{-}$)
have charge ($-1$, $-1$, $1$, $1$), the supersymmetry parameters ($\epsilon_{+}$, $\epsilon_{-}$, $\tilde{\epsilon}_{+}$, $\tilde{\epsilon}_{-}$)
have charge ($1$, $1$, $-1$, $-1$), the vector multiplet component fields
($A_{\mu}$, $\sigma$, $\tilde{\sigma}$, $D_{E}$, $\lambda_{+}$, $\lambda_{-}$, $\tilde{\lambda}_{+}$, $\tilde{\lambda}_{-}$)
have charge ($0$, $0$, $0$, $0$, $1$, $1$, $-1$, $-1$), and the chiral multiplet component
fields ($\phi$, $\tilde{\phi}$, $\psi_{+}$, $\psi_{-}$, $\tilde{\psi}_{+}$, $\tilde{\psi}_{-}$, $F$, $\tilde{F}$)
have charge ($q_{V}$, $-q_{V}$, $q_{V}-1$, $q_{V}-1$, $1-q_{V}$, $1-q_{V}$, $q_{V}-2$, $2-q_{V}$).

\subsection{$A$-twisted $\mathcal{N}=(2,2)$ supersymmetry on $S^{2}$\label{subsec:twisted--supersymmetry}}

This section reviews the passage from $\mathcal{N}=(2,2)$ supersymmetry
on $\mathbb{R}^{2}$ to the $A$-model on $S^{2}$ via the topological
$A$-twist \cite{Witten:1988xj,Closset:2014pda,Benini:2016qnm}. We
first describe the $A$-twist, then detail the field content and supersymmetry
transformations of the twisted theory. 

The metric, vielbein, and spin connection of the $S^{2}$ are 
\begin{equation}
\text{d}s^{2}=R^{2}\left(\text{d}\theta^{2}+\sin^{2}\theta\,\text{d}\phi^{2}\right),\quad e^{1}=R\text{d}\theta,\quad e^{2}=R\sin\theta\text{d}\phi,\quad\omega=\cos\theta\text{d}\phi,
\end{equation}
where $R$ is the radius of the $S^{2}$, $\phi$ is $2\pi$ periodic,
and $0\leq\theta\leq\pi$. We further define the holomorphic vielbein
$e^{z}=e^{1}+ie^{2}$, and denote the volume form on the $S^{2}$
by $\text{d}\mu$. The $A$-twist is a solution of the Killing spinor
equations of the $S^{2}$ that preserves half of the $\mathcal{N}=\left(2,2\right)$
supercharges of $\mathbb{R}^{2}$. The solution is \cite{Benini:2016qnm}
\begin{equation}
A^{V}_{\mu}=\frac{1}{2}\omega_{\mu},\qquad\epsilon=\begin{pmatrix}0\\
\epsilon_{-}
\end{pmatrix},\qquad\tilde{\epsilon}=\begin{pmatrix}\tilde{\epsilon}_{+}\\
0
\end{pmatrix},\qquad\mathcal{H}=0,\qquad\widetilde{\mathcal{H}}=0,
\end{equation}
where $A^{V}_{\mu}$ is a connection of a background $\text{U}(1)_{V}$
symmetry, $\omega_{\mu}$ is the spin connection on $S^{2}$, $\partial_{\mu}\epsilon_{-}=\partial_{\mu}\tilde{\epsilon}_{+}=0$,
and $\mathcal{H},\widetilde{\mathcal{H}}$ are bosonic scalars.

Under the $A$-twisted rotation group $\text{U}(1)^{\prime}_{E}$,
the supercharges ($Q_{+}$, $Q_{-}$, $\tilde{Q}_{+}$, $\tilde{Q}_{-}$) have
charge ($0$, $-2$, $2$, $0$), the supersymmetry parameters
($\epsilon_{+}$, $\epsilon_{-}$, $\tilde{\epsilon}_{+}$, $\tilde{\epsilon}_{-}$)
have charge ($2$, $0$, $0$, $-2$), the vector multiplet component fields
($A_{z}$, $A_{\bar{z}}$, $\sigma$, $\tilde{\sigma}$, $D_{E}$, $\lambda_{+}$, $\lambda_{-}$, $\tilde{\lambda}_{+}$, $\tilde{\lambda}_{-}$)
have charge ($2$, $-2$, $0$, $0$, $0$, $2$, $0$, $0$, $-2$), and the chiral multiplet component fields
($\phi$, $\tilde{\phi}$, $\psi_{+}$, $\psi_{-}$, $\tilde{\psi}_{+}$, $\tilde{\psi}_{-}$, $F$, $\tilde{F}$)
have charge ($q_{V}$, $-q_{V}$, $q_{V}$, $q_{V}-2$, $2-q_{V}$, $-q_{V}$, $q_{V}-2$, $2-q_{V}$).
Quantities with $\text{U}(1)^{\prime}_{E}$ charge $0$ are scalars,
those with charge $2$ are holomorphic vectors, and those with charge
$-2$ are anti-holomorphic vectors. To make the $A$-twisted spins
manifest, we set $q_{V}=0$ and relabel
\begin{equation}
\begin{array}{llll}
Q:=Q_{+}, & \widetilde{Q}:=\widetilde{Q}_{-}, & \epsilon:=\epsilon_{-}, & \tilde{\epsilon}:=\tilde{\epsilon}_{+},\\
\lambda_{z}:=\lambda_{+}, & \lambda:=\lambda_{-}, & \tilde{\lambda}:=\tilde{\lambda}_{+}, & \lambda_{\bar{z}}:=\tilde{\lambda}_{-}\\
\psi:=\psi_{+}, & \psi_{\bar{z}}:=\psi_{-}, & \psi_{z}:=\tilde{\psi}_{+}, & \tilde{\psi}:=\tilde{\psi}_{-}\\
F_{\bar{z}}:=F, & F_{z}:=\tilde{F}.
\end{array}
\end{equation}
The scalar supercharge of the $A$-model 
\begin{equation}
Q_{A}:=Q+\widetilde{Q},
\end{equation}
 acts on the fields of the vector multiplet as 
\begin{equation}
\begin{array}{l}
Q_{A}A_{z}=i\lambda_{z},\\
Q_{A}A_{\bar{z}}=i\tilde{\lambda}_{\bar{z}},\\
Q_{A}\sigma=0,\\
Q_{A}\tilde{\sigma}=2\left(\lambda-\tilde{\lambda}\right),\\
Q_{A}D_{E}=-2i\left(D_{z}\tilde{\lambda}_{\bar{z}}-D_{\bar{z}}\lambda_{z}\right)-i[\sigma,\tilde{\lambda}+\lambda],\\
Q_{A}\lambda_{z}=-2iD_{z}\sigma,\\
Q_{A}\lambda=-D_{E}-i\star F+\frac{i}{2}[\sigma,\tilde{\sigma}],\\
Q_{A}\tilde{\lambda}=-D_{E}-i\star F-\frac{i}{2}[\sigma,\tilde{\sigma}],\\
Q_{A}\tilde{\lambda}_{\bar{z}}=-2iD_{\bar{z}}\sigma,\\
Q_{A}\star F=2D_{z}\tilde{\lambda}_{\bar{z}}-2D_{\bar{z}}\lambda_{z},
\end{array}\label{eq:susyS2_vpletVariation2}
\end{equation}
where the scalar $\star F=F_{12}=-2iF_{z\bar{z}}$ is the Hodge dual
of the field strength. The scalar supercharge of the \emph{$A$}-model
acts on the chiral multiplet as
\begin{equation}
\begin{array}{l}
Q_{A}\phi=\sqrt{2}\psi,\\
Q_{A}\tilde{\phi}=-\sqrt{2}\tilde{\psi},\\
Q_{A}\psi=i\sqrt{2}\sigma\phi,\\
Q_{A}\psi_{\bar{z}}=2i\sqrt{2}D_{\bar{z}}\phi-i\sqrt{2}F_{\bar{z}},\\
Q_{A}\psi_{z}=2i\sqrt{2}D_{z}\tilde{\phi}+i\sqrt{2}F_{z},\\
Q_{A}\tilde{\psi}=i\sqrt{2}\sigma\tilde{\phi},\\
Q_{A}F_{\bar{z}}=2\sqrt{2}D_{\bar{z}}\psi+2\tilde{\lambda}_{\bar{z}}\phi-\sqrt{2}\sigma\psi_{\bar{z}},\\
Q_{A}F_{z}=2\sqrt{2}D_{z}\tilde{\psi}+2\tilde{\phi}\lambda_{z}-\sqrt{2}\sigma\psi_{z}.
\end{array}\label{eq:susyS2_chpletVariation2}
\end{equation}
The transformations (\ref{eq:susyS2_vpletVariation2}) and (\ref{eq:susyS2_chpletVariation2})
are obtained by setting $\epsilon_{+}=\tilde{\epsilon}_{-}=0$ and
$\epsilon_{-}=\tilde{\epsilon}_{+}=1$ in (\ref{eq:susyR2_vpletVariation1})
and (\ref{eq:susyR2_chpletVariation1}), respectively, and relabeling
the fields. By defining the auxiliary fields \cite{Witten:1993xi,Ohta:2019odi}
\begin{equation}
\begin{array}{l}
H=D_{E}+i\star F,\\
H_{\bar{z}}=-iF_{\bar{z}}+2iD_{\bar{z}}\phi,\\
H_{z}=-iF_{z}-2iD_{z}\tilde{\phi},
\end{array}
\end{equation}
the equations describing the action of $Q_{A}$ on $\lambda,\tilde{\lambda},D_{E},\psi_{\bar{z}},\psi_{z},F_{\bar{z}},F_{z}$
in (\ref{eq:susyS2_vpletVariation2}) and (\ref{eq:susyS2_chpletVariation2})
become 
\begin{equation}
\begin{array}{cll}
Q_{A}H=-i[\sigma,\lambda+\tilde{\lambda}], & \quad Q_{A}\psi_{\bar{z}}=\sqrt{2}H_{\bar{z}}, & \quad Q_{A}\psi_{z}=-\sqrt{2}H_{z},\\
Q_{A}\lambda=-H+\frac{i}{2}[\sigma,\tilde{\sigma}], & \quad Q_{A}H_{\bar{z}}=i\sqrt{2}\sigma\psi_{\bar{z}}, & \quad Q_{A}H_{z}=i\sqrt{2}\sigma\psi_{z}.\\
Q_{A}\tilde{\lambda}=-H-\frac{i}{2}[\sigma,\tilde{\sigma}],
\end{array}
\end{equation}
The algebra of $Q_{A}$ is
\begin{equation}
\begin{array}{ccl}
Q^{2}_{A}A_{\mu} & = & 2D_{\mu}\sigma,\\
Q^{2}_{A}\sigma & = & 0,\\
Q^{2}_{A}X & = & 2i\left[\sigma,X\right],
\end{array}
\end{equation}
where $X$ denotes the fields of the vector multiplet, $\tilde{\sigma},H,\lambda_{\mu},\lambda,\tilde{\lambda}$,
as well as the chiral multiplet, $\phi,\tilde{\phi},\psi,\tilde{\psi},\psi_{\mu},H_{\mu}$.
Since $Q_{A}$ squares to a gauge transformation with parameter $\vartheta=2\sigma$,
it is nilpotent when acting on gauge-invariant functionals of the
fields of the vector and chiral multiplets.

\section{Localization \label{sec:Localization}}

In this section, we use localization techniques to evaluate path integrals
of $A$-twisted $\mathcal{N}=(2,2)$ theories of vector and chiral
multiplets on $S^{2}$, focusing on the chiral multiplet one-loop
contribution. We construct the localizing terms, determine the localization
locus, then expand the localizing term in a functional Taylor series
around the locus configurations in the localizing limit. The fluctuation
operators are determined by formulating a matrix expression for the
localizing term to quadratic order in fluctuations. The one-loop contribution
is computed by expanding the fluctuations in a basis of monopole spherical
harmonics. In evaluating the integrals over fluctuations, we identify
a bosonic scalar mode with purely imaginary eigenvalue, associated
to an oscillatory integral lacking a  positive definite quadratic
form. The oscillatory integral is evaluated as a distribution, while
the integrals over the remaining fluctuations are evaluated as a ratio
of determinants. Finally, we collect the results of localization,
and present an exact formula in which observables of abelian $A$-twisted
$\mathcal{N}=(2,2)$ GLSMs on $S^{2}$ are described by a distribution
integrated along a real contour.

\subsection{Localizing term \label{subsec:localizing-term}}

The localizing term is the action functional that is used to deform
then localize the original path integral. It is constructed by defining
fermionic functionals of the fields, then acting on them with the
localizing supercharge $Q_{A}$. 

For the vector multiplet, we define the functional 
\begin{equation}
V_{\text{vec}}(t,\tau)=-\frac{1}{4}\int\text{d}\mu\,\text{Tr}\,\left((\tilde{\lambda}+\lambda)(H-2i\star F)-2i(\lambda_{z}D_{\bar{z}}\tilde{\sigma}+\lambda_{\bar{z}}D_{z}\tilde{\sigma})-\frac{i\tau}{2}(\tilde{\lambda}-\lambda)[\sigma,\tilde{\sigma}]+2t(\tilde{\lambda}+\lambda)\tilde{\sigma}\right),\label{eq:locTerm1_Vvec}
\end{equation}
where $t$ and $\tau$ are real parameters. Acting with $Q_{A}$ on
(\ref{eq:locTerm1_Vvec}), we obtain 
\begin{eqnarray}
Q_{A}V_{\text{vec}}(t,\tau) & = & \int\text{d}\mu\,\text{Tr}\,\Bigg(\frac{1}{2}\left(\star F+it\tilde{\sigma}\right)^{2}+\frac{1}{2}\left(H-i\star F+t\tilde{\sigma}\right)^{2}+\frac{1}{2}D_{\mu}\tilde{\sigma}D^{\mu}\sigma\label{eq:locTerm2_QVvec}\\
 &  & \qquad\qquad+2i\lambda_{z}D_{\bar{z}}\tilde{\lambda}-2i\lambda_{\bar{z}}D_{z}\lambda+i\lambda_{\bar{z}}\left[\tilde{\sigma},\lambda_{z}\right]-\frac{i}{2}\tilde{\lambda}[\sigma,\lambda]-\frac{i}{4}\tilde{\lambda}[\sigma,\tilde{\lambda}]-\frac{i}{4}\lambda[\sigma,\lambda]\nonumber \\
 &  & \qquad\qquad+2t\tilde{\lambda}\lambda+\tau\left(\frac{1}{8}\left[\sigma,\tilde{\sigma}\right]^{2}-\frac{i}{2}\tilde{\lambda}[\sigma,\lambda]+\frac{i}{4}\tilde{\lambda}[\sigma,\tilde{\lambda}]+\frac{i}{4}\lambda[\sigma,\lambda]\right)\Bigg),\nonumber 
\end{eqnarray}
For $t=0$ and $\tau=1$, this is the standard D-term Yang-Mills action
for the $A$-model vector multiplet, which served as the localizing
term in \cite{Benini:2015noa,Closset:2015rna,Benini:2016hjo,Ohta:2019odi}.
For $t>0$ and $\tau=0$, (\ref{eq:locTerm2_QVvec}) is a $Q_{A}$-exact
deformation of the standard D-term action, which played the role of
the localizing term in \cite{Leeb-Lundberg:2023jsj,Witten:1992xu}.
The $t$-dependent deformation may be regarded as choosing a quadratic
$Q_{A}$-exact superpotential in the F-term action of the vector multiplet.
We proceed with real $t>0$ and $\tau=0$. Observe that for real $t>0$,
there is a mass term for $\lambda$ and $\tilde{\lambda}$, as well
as a mixing term between $\star F$ and $\tilde{\sigma}$. 

For the chiral multiplet, we define the functional
\begin{equation}
V_{\text{chi}}=\frac{i}{2\sqrt{2}}\int\text{d}\mu\,\left(F_{z}\psi_{\bar{z}}-\psi_{z}F_{\bar{z}}+2(\tilde{\phi}D_{z}\psi_{\bar{z}}-\psi_{z}D_{\bar{z}}\phi)+\sqrt{2}\tilde{\phi}(\tilde{\lambda}+\lambda)\phi-\tilde{\phi}\tilde{\sigma}\psi-\tilde{\psi}\tilde{\sigma}\phi\right).\label{eq:locTerm3_Vchi}
\end{equation}
Acting with $Q_{A}$ on (\ref{eq:locTerm3_Vchi}), we obtain 
\begin{eqnarray}
Q_{A}V_{\text{chi}} & = & \int\text{d}\mu\,\big(D_{\mu}\tilde{\phi}D^{\mu}\phi-i\tilde{\phi}(H-i\star F)\phi+\frac{1}{2}\tilde{\phi}\{\sigma,\tilde{\sigma}\}\phi+F_{z}F_{\bar{z}}\label{eq:locTerm4_QVchi}\\
 &  & \qquad\quad+2i\psi_{z}D_{\bar{z}}\psi-2i\tilde{\psi}D_{z}\psi_{\bar{z}}-i\psi_{z}\sigma\psi_{\bar{z}}+i\tilde{\psi}\tilde{\sigma}\psi\nonumber \\
 &  & \qquad\quad+i\sqrt{2}\tilde{\phi}\lambda_{z}\psi_{\bar{z}}+i\sqrt{2}\psi_{z}\tilde{\lambda}_{\bar{z}}\phi-i\sqrt{2}\tilde{\phi}\lambda\psi-i\sqrt{2}\tilde{\psi}\tilde{\lambda}\phi\,\big),\nonumber 
\end{eqnarray}
where $D_{E}=H-i\star F$. This is the standard D-term matter action
for the $A$-model chiral multiplet, which served as the localizing
term in \cite{Benini:2015noa,Closset:2015rna,Benini:2016hjo,Ohta:2019odi}. 

In what follows, we consider localization of the action functional
\begin{equation}
S_{\text{loc}}=\frac{1}{g^{2}}Q_{A}V_{\text{vec}}(t>0,\tau=0)+\frac{1}{h^{2}}Q_{A}V_{\text{chi}},\label{eq:locus0_QV}
\end{equation}
in the limit $g,h\to0$.

\subsection{Localization locus \label{subsec:localization-locus}}

The localization locus is the space of field configurations for which
the localizing term vanishes along a particular integration contour. The
Lie algebra of the gauge group $\mathfrak{g}=\text{Lie}G$ is taken
to consist of hermitian matrices such that the positive definite metric
on $\mathfrak{g}$ is $\text{Tr}(a,b)$. The fields in the vector
multiplet are valued in the adjoint representation of $\mathfrak{g}$,
while the fields in the chiral multiplet are valued in a representation
$\mathcal{R}$ of $\mathfrak{g}$. The fields are generically complex
in euclidean signature. An integration contour is chosen by specifying
reality conditions for the fields and any contour for which the path
integral converges is acceptable. 

The zero-action bosonic field configurations are specified by the
constraint 
\begin{equation}
\begin{array}{ccl}
0 & = & \frac{1}{2g^{2}}\text{Tr}\left(\left(\star F+it\tilde{\sigma}\right)^{2}+\left(H-i\star F+t\tilde{\sigma}\right)^{2}+D_{\mu}\tilde{\sigma}D^{\mu}\sigma\right)\\
 &  & +\frac{1}{h^{2}}\left(D_{\mu}\tilde{\phi}D^{\mu}\phi-i\tilde{\phi}(H-i\star F)\phi+\frac{1}{2}\tilde{\phi}\{\sigma,\tilde{\sigma}\}\phi+F_{z}F_{\bar{z}}\right),
\end{array}\label{eq:locus1_ZeroActionConfigurations}
\end{equation}
while the BPS equations are 
\begin{equation}
\begin{array}{lclcccl}
0 & = & -2iD_{z}\sigma, &  & 0 & = & i\sqrt{2}\sigma\phi,\\
0 & = & -H+\frac{i}{2}[\sigma,\tilde{\sigma}], &  & 0 & = & 2i\sqrt{2}D_{\bar{z}}\phi-i\sqrt{2}F_{\bar{z}},\\
0 & = & -H-\frac{i}{2}[\sigma,\tilde{\sigma}], &  & 0 & = & 2i\sqrt{2}D_{z}\tilde{\phi}+i\sqrt{2}F_{z},\\
0 & = & -2iD_{\bar{z}}\sigma, &  & 0 & = & i\sqrt{2}\sigma\tilde{\phi},
\end{array}\label{eq:locus2_bpsEqs}
\end{equation}
where $H=D_{E}+i\star F.$

Let us begin with the fields in the chiral multiplet. We choose the
reality conditions $\phi=\tilde{\phi}^{\dag}$ and $F_{z}=F^{\dag}_{\bar{z}}$
such that $D_{\mu}\tilde{\phi}D^{\mu}\phi=|D_{\mu}\phi|^{2}$, $F_{z}F_{\bar{z}}=|F_{\mu}|^{2}$,
and so on. The field configurations 
\begin{equation}
\phi=0,\quad F_{\mu}=0,\label{eq:locus3_moduliChplet}
\end{equation}
are zero-action in the sense that they set the second line in (\ref{eq:locus1_ZeroActionConfigurations})
to zero, and they are BPS configurations in the sense that they solve
the equations in the second column of (\ref{eq:locus2_bpsEqs}). 

Next we consider the fields in the vector multiplet. The auxiliary
field $H$ is integrated out by setting it to its on-shell value in
the action. Taking real $\sigma$ and purely imaginary $\tilde{\sigma}=i\tilde{\sigma}_{E}$,
the condition for the first term in (\ref{eq:locus1_ZeroActionConfigurations})
to vanish is 
\begin{equation}
0=\text{Tr}\,(\star F-t\tilde{\sigma}_{E})^{2}+i\text{Tr}\,D_{\mu}\tilde{\sigma}_{E}D^{\mu}\sigma.\label{eq:locus4_ZeroActionVplet}
\end{equation}
By requiring the integrals over $\sigma$ and $\tilde{\sigma}_{E}$
to have stationary phase, (\ref{eq:locus4_ZeroActionVplet}) reduces
to 
\begin{equation}
\star F=t\tilde{\sigma}_{E},\quad D_{\mu}\tilde{\sigma}_{E}=0,\quad D_{\mu}\sigma=0.\label{eq:locus5_ZeroActionVplet}
\end{equation}
The first two equations of (\ref{eq:locus5_ZeroActionVplet}) imply
the Yang-Mills equations on the two-sphere, 
\begin{equation}
\frac{1}{t}D_{\mu}\star F=0,\label{eq:locus6_YMeq}
\end{equation}
for real $t>0$. The solutions of (\ref{eq:locus6_YMeq}) are either
flat connections or Yang-Mills connections \cite{Atiyah:1982fa}.
The flux of a Yang-Mills connection on $S^{2}$ is GNO quantized \cite{GODDARD19771}
\begin{equation}
\frac{1}{2\pi}\int_{S^{2}}F=\mathfrak{m}\in\Lambda^{G}_{\text{cochar}},\label{eq:locus7_quantization}
\end{equation}
where $\Lambda^{G}_{\text{cochar}}=\{\,\gamma\in\mathfrak{h}\,\vert\,e^{2\pi i\gamma}=1_{G}\,\}$
is the cocharacter lattice of $G$, $\mathfrak{h}$ is the Cartan
subalgebra of $\mathfrak{g}$, and $1_{G}$ is the identity element
in $G$. Consequently, the solutions of the first two equations in
(\ref{eq:locus5_ZeroActionVplet}) are parameterized by the quantized
flux $\mathfrak{\mathfrak{m}}$. 

The final equation in (\ref{eq:locus5_ZeroActionVplet}) has two types
of solutions depending on whether the connection $A$ is a reducible
or irreducible solution of the Yang-Mills equations (\ref{eq:locus6_YMeq}).
For a reference, see e.g. section 14.4 in \cite{Deligne:1999qp}.
We take $A$ to be a reducible Yang-Mills connection, such that the
solutions of $D_{\mu}\sigma=0$ are non-zero configurations $\sigma$
that are constant ($\partial_{\mu}\sigma=0$) and commuting $([A_{\mu},\sigma]=0)$.
These solutions are denoted $u$. Acting on the final equation in
(\ref{eq:locus5_ZeroActionVplet}) with covariant derivatives, we
find $[F_{\mu\nu},\sigma]=0$, which implies that $\sigma$ can be
conjugated into $\mathfrak{h}$ using an element of $G$, and that
$[\sigma,\tilde{\sigma}]=0$. 

To summarize, the solutions of (\ref{eq:locus5_ZeroActionVplet})
and (\ref{eq:locus2_bpsEqs}) are 
\begin{equation}
\star F=\frac{\mathfrak{\mathfrak{m}}}{2R^{2}},\quad\tilde{\sigma}=\frac{i\mathfrak{\mathfrak{m}}}{2tR^{2}},\quad\sigma=u,\quad H=0,\quad\phi=0,\quad F_{\mu}=0,\label{eq:locus8_moduliVplet}
\end{equation}
where $u\in\mathfrak{h}$ is a continuous real modulus\footnote{Note that our $u$ differs from the one in \cite{Leeb-Lundberg:2023jsj}
as $u^{\text{here}}=-u^{\text{there}}/2$.} parameterizing $\sigma$, and $\mathfrak{\mathfrak{m}}\in\Lambda^{G}_{\text{cochar}}\subset\mathfrak{h}$
is a discrete modulus parameterizing $A_{\mu}$ and $\tilde{\sigma}$. 

Let us remark on the absence of fermionic zero modes. The fermionic
scalars in the vector multiplet $\lambda,\tilde{\lambda}$ do not
have zero modes since they are lifted for real $t>0$ \cite{Leeb-Lundberg:2023jsj}.
The fermionic vector fields $\lambda_{\mu},\psi_{\mu}$ do not have
zero modes at genus $g=0$, but these can appear at $g>0$. While
one might be concerned about zero modes for the fermionic scalars
in the chiral multiplet, $\psi,\tilde{\psi}$, our one-loop analysis
shows that these do not appear. Specifically, in the fluctuation operator
of the chiral multiplet, the eigenvalues associated to $\psi,\tilde{\psi}$
are non-zero for generic non-zero $\mathfrak{\mathfrak{m}}$ and $u$.
Notice that this locus differs from those considered in \cite{Benini:2015noa,Closset:2015rna,Benini:2016hjo,Ohta:2019odi},
where zero modes of fermionic scalars play a significant role.

\subsection{One-loop contribution\label{subsec:One-loop-contribution}}

In this section, we evaluate the one-loop contribution of the $A$-twisted
$\mathcal{N}=(2,2)$ chiral multiplet on $S^{2}$. We begin by expanding
the localizing term $h^{-2}Q_{A}V_{\text{chi }}$ in a functional
Taylor series around the locus configurations in the limit $h\to0$,
then determining the fluctuation operators. The integrals over fluctuating
modes are evaluated using monopole spherical harmonics on $S^{2}$.
In the mode analysis, we identify a fluctuating mode of the bosonic
scalar with purely imaginary eigenvalue, corresponding to an oscillatory
integral without a positive definite quadratic form. We retain the
oscillatory integral, and the one-loop contribution is a product of
the oscillatory integral and a ratio of fluctuation determinants.
Following this, we specialize the one-loop contribution to the abelian
case and reduce the oscillatory integral to a linear combination of
distributions. Finally, we collect the results.

To begin, the fields in the localizing term (\ref{eq:locus0_QV})
are expressed as $X=X_{0}+hX^{\prime}$, where $X$ denotes the fields
of the chiral and vector multiplets, $X_{0}$ denotes zero modes,
and $X^{\prime}$ denotes fluctuations. The zero modes, described
in (\ref{eq:locus3_moduliChplet}) and (\ref{eq:locus8_moduliVplet}),
are zero-action BPS configurations that solve (\ref{eq:locus1_ZeroActionConfigurations})
and (\ref{eq:locus2_bpsEqs}). Specifically, the fields of the chiral
multiplet are expressed as 
\begin{equation}
\begin{array}{ccc}
\phi=h\phi^{\prime}, &  & \tilde{\phi}=h\tilde{\phi}^{\prime},\\
F_{\overline{z}}=hF^{\prime}_{\overline{z}}, &  & F_{z}=hF^{\prime}_{z},\\
\psi=h\psi^{\prime}, &  & \tilde{\psi}=h\tilde{\psi}^{\prime},\\
\psi_{\overline{z}}=h\psi^{\prime}_{\overline{z}}, &  & \psi_{z}=h\psi^{\prime}_{z},
\end{array}\label{eq:expansion1_fields}
\end{equation}
where the zero modes are all zero and only fluctuations remain. Consequently,
all the terms in the functional Taylor series expansion of $\frac{1}{h^{2}}Q_{A}V_{\text{chi}}$
vanish in the $h\to0$ limit, except for the term at quadratic order
in fluctuations. The term that survives at quadratic order in fluctuations
is 
\begin{equation}
\begin{array}{ccl}
\tilde{S}_{\text{chi}} & = & \int\text{d}\mu\,\big(\,D^{0}_{\mu}\tilde{\phi}^{\prime}D^{\mu}_{0}\phi^{\prime}-i\tilde{\phi}^{\prime}(H_{0}-i\star F_{0})\phi^{\prime}+\frac{1}{2}\tilde{\phi}^{\prime}\{\sigma_{0},\tilde{\sigma}_{0}\}\phi^{\prime}+F^{\prime}_{z}F^{\prime}_{\bar{z}}\\
 &  & \qquad\quad+2i\psi^{\prime}_{z}D^{0}_{\bar{z}}\psi^{\prime}-2i\tilde{\psi}^{\prime}D^{0}_{z}\psi^{\prime}_{\overline{z}}-i\psi^{\prime}_{z}\sigma_{0}\psi^{\prime}_{\overline{z}}+i\tilde{\psi}^{\prime}\tilde{\sigma}_{0}\psi^{\prime}\\
 &  & \qquad\quad+i\sqrt{2}\tilde{\phi}^{\prime}\lambda^{0}_{z}\psi^{\prime}_{\overline{z}}+i\sqrt{2}\psi^{\prime}_{z}\tilde{\lambda}^{0}_{\bar{z}}\phi^{\prime}-i\sqrt{2}\tilde{\phi}^{\prime}\lambda_{0}\psi^{\prime}-i\sqrt{2}\tilde{\psi}^{\prime}\tilde{\lambda}_{0}\phi^{\prime}\,\big),
\end{array}\label{eq:expansion2_SquadAction}
\end{equation}
where $D^{0}_{\mu}=\partial_{\mu}-iA^{0}_{\mu}$. The quadratic action
(\ref{eq:expansion2_SquadAction}) can be expressed as 
\begin{equation}
\tilde{S}_{\text{chi}}=\int\text{d}\mu\left(\bar{\Phi}\Delta_{\mathrm{B}}\Phi+\bar{\Psi}\Delta_{\mathrm{F}}\Psi\right),\label{eq:expansion2_SquadMatrix}
\end{equation}
where the bosonic and fermionic fluctuation operators are respectively
\begin{eqnarray}
\bar{\Phi}\Delta_{\mathrm{B}}\Phi & = & \frac{1}{2}(\phi^{\prime},\tilde{\phi}^{\prime},F^{\prime}_{z},F^{\prime}_{\bar{z}})\left(\begin{array}{cccc}
-D^{0}_{\mu}D^{\mu}_{0}+iH_{0}+\star F_{0}+\frac{\{\sigma_{0},\tilde{\sigma}_{0}\}}{2} & 0 & 0 & 0\\
0 & -D^{0}_{\mu}D^{\mu}_{0}-iH_{0}-\star F_{0}+\frac{\{\sigma_{0},\tilde{\sigma}_{0}\}}{2} & 0 & 0\\
0 & 0 & 1 & 0\\
0 & 0 & 0 & 1
\end{array}\right)\left(\begin{array}{c}
\tilde{\phi}^{\prime}\\
\phi^{\prime}\\
F^{\prime}_{\bar{z}}\\
F^{\prime}_{z}
\end{array}\right),\label{eq:expansion3_Obos}\\
\bar{\Psi}\Delta_{\mathrm{F}}\Psi & = & \frac{1}{2}(\psi^{\prime},\tilde{\psi}^{\prime},\psi^{\prime}_{z},\psi^{\prime}_{\bar{z}})\left(\begin{array}{cccc}
i\tilde{\sigma}_{0} & 0 & 0 & 2iD^{0}_{\bar{z}}\\
0 & i\tilde{\sigma}_{0} & -2iD^{0}_{z} & 0\\
0 & 2iD^{0}_{\bar{z}} & -i\sigma_{0} & 0\\
-2iD^{0}_{z} & 0 & 0 & -i\sigma_{0}
\end{array}\right)\left(\begin{array}{c}
\tilde{\psi}^{\prime}\\
\psi^{\prime}\\
\psi^{\prime}_{\bar{z}}\\
\psi^{\prime}_{z}
\end{array}\right).\label{eq:expansion4_Ofer}
\end{eqnarray}
The fluctuating modes are decomposed in a basis of weights of representations
$\rho\in\mathcal{R}$ of the gauge group, as well as a basis of monopole
spherical harmonics on $S^{2}$, which reads
\begin{eqnarray}
\phi^{\prime}=\sum_{\rho\in\mathcal{R}}\sum^{\infty}_{j=0}\sum^{j}_{j_{3}=-j}\phi^{\rho}_{j,j_{3}}Y^{-\frac{\rho(\mathfrak{m})}{2}}_{j,j_{3}}e_{\rho}, &  & \tilde{\phi}^{\prime}=\sum_{\rho\in\mathcal{R}}\sum^{\infty}_{j^{\prime}=0}\sum^{j^{\prime}}_{j^{\prime}_{3}=-j^{\prime}}\overline{\phi}^{\rho}_{j^{\prime},j^{\prime}_{3}}Y^{-\frac{\rho(\mathfrak{m})}{2}}_{j^{\prime},j^{\prime}_{3}}\left(e_{\rho}\right)^{\dag},\label{eq:expansion5_basisFields1}\\
F^{\prime}_{\overline{z}}=\sum_{\rho\in\mathcal{R}}\sum^{\infty}_{j=0}\sum^{j}_{j_{3}=-j}f^{\rho}_{j,j_{3}}Y^{-\frac{\rho(\mathfrak{m})}{2}-1}_{j,j_{3}}e_{\rho}, &  & F^{\prime}_{z}=\sum_{\rho\in\mathcal{R}}\sum^{\infty}_{j^{\prime}=0}\sum^{j^{\prime}}_{j^{\prime}_{3}=-j^{\prime}}\overline{f}^{\rho}_{j^{\prime},j^{\prime}_{3}}Y^{-\frac{\rho(\mathfrak{m})}{2}+1}_{j^{\prime},j^{\prime}_{3}}\left(e_{\rho}\right)^{\dag},\label{eq:expansion5_basisFields2}\\
\psi^{\prime}=\sum_{\rho\in\mathcal{R}}\sum^{\infty}_{j=0}\sum^{j}_{j_{3}=-j}\psi^{\rho}_{j,j_{3}}Y^{-\frac{\rho(\mathfrak{m})}{2}}_{j,j_{3}}e_{\rho}, &  & \tilde{\psi}^{\prime}=\sum_{\rho\in\mathcal{R}}\sum^{\infty}_{j^{\prime}=0}\sum^{j^{\prime}}_{j^{\prime}_{3}=-j^{\prime}}\overline{\psi}^{\rho}_{j^{\prime},j^{\prime}_{3}}Y^{-\frac{\rho(\mathfrak{m})}{2}}_{j^{\prime},j^{\prime}_{3}}\left(e_{\rho}\right)^{\dag},\label{eq:expansion5_basisFields3}\\
\psi^{\prime}_{\overline{z}}=\sum_{\rho\in\mathcal{R}}\sum^{\infty}_{j=0}\sum^{j}_{j_{3}=-j}\zeta^{\rho}_{j,j_{3}}Y^{-\frac{\rho(\mathfrak{m})}{2}-1}_{j,j_{3}}e_{\rho}, &  & \psi^{\prime}_{z}=\sum_{\rho\in\mathcal{R}}\sum^{\infty}_{j^{\prime}=0}\sum^{j^{\prime}}_{j^{\prime}_{3}=-j^{\prime}}\overline{\zeta}^{\rho}_{j^{\prime},j^{\prime}_{3}}Y^{-\frac{\rho(\mathfrak{m})}{2}+1}_{j^{\prime},j^{\prime}_{3}}\left(e_{\rho}\right)^{\dag}.\label{eq:expansion5_basisFields4}
\end{eqnarray}
where $\dagger$ denotes hermitian conjugation and the bar denotes
complex conjugation. The basis of weights $e_{\rho}$ obeys $(e_{\rho^{\prime}})^{\dag}e_{\rho}=1$,
where $\rho,\rho^{\prime}$ are weights of a representation $\mathcal{R}$
of the Lie algebra of the gauge group $\mathfrak{g}=\text{Lie}G$.
The monopole spherical harmonics, described in \cite{Benini:2012ui,Benini:2015noa},
are 
\begin{equation}
Y^{s}_{j,j_{3}}\text{ for }j\geq\left|s\right|,\qquad Y^{s+1}_{j,j_{3}}\text{ for }j\geq\left|s+1\right|,\qquad Y^{s-1}_{j,j_{3}}\text{ for }j\geq\left|s-1\right|,\label{eq:harmonics}
\end{equation}
where $s=-\frac{1}{2}c_{1}$ is the effective spin and $c_{1}=\rho(\mathfrak{m})$
is the first Chern number of the line bundle. The harmonics obey 
\begin{equation}
\begin{array}{ccccl}
D^{0}_{z}Y^{s}_{j,j_{3}}=\frac{s_{+}}{2R}Y^{s+1}_{j,j_{3}}, &  & D^{0}_{z}D^{0}_{\bar{z}}Y^{s+1}_{j,j_{3}}=-\frac{s^{2}_{+}}{4R^{2}}Y^{s+1}_{j,j_{3}}, &  & D^{0}_{z}D^{0}_{z}Y^{s-1}_{j,j_{3}}=\frac{s_{-}s_{+}}{4R^{2}}Y^{s+1}_{j,j_{3}},\\
\\D^{0}_{\bar{z}}Y^{s}_{j,j_{3}}=-\frac{s_{-}}{2R}Y^{s-1}_{j,j_{3}}, &  & D^{0}_{\bar{z}}D^{0}_{z}Y^{s-1}_{j,j_{3}}=-\frac{s^{2}_{-}}{4R^{2}}Y^{s-1}_{j,j_{3}}, &  & D^{0}_{\bar{z}}D^{0}_{\bar{z}}Y^{s+1}_{j,j_{3}}=\frac{s_{-}s_{+}}{4R^{2}}Y^{s-1}_{j,j_{3}},\\
\\D^{0}_{\bar{z}}Y^{s+1}_{j,j_{3}}=-\frac{s_{+}}{2R}Y^{s}_{j,j_{3}}, &  & D^{0}_{z}D^{0}_{\bar{z}}Y^{s}_{j,j_{3}}=-\frac{s^{2}_{-}}{4R^{2}}Y^{s}_{j,j_{3}}, &  & D^{0}_{z}Y^{s+1}_{j,j_{3}}=0,\\
\\D^{0}_{z}Y^{s-1}_{j,j_{3}}=\frac{s_{-}}{2R}Y^{s}_{j,j_{3}}, &  & D^{0}_{\bar{z}}D^{0}_{z}Y^{s}_{j,j_{3}}=-\frac{s^{2}_{+}}{4R^{2}}Y^{s}_{j,j_{3}}, &  & D^{0}_{\bar{z}}Y^{s-1}_{j,j_{3}}=0,
\end{array}\label{eq:eigenvalues}
\end{equation}
where $s_{\pm}=\sqrt{j(j+1)-s(s\pm1)}$. 

The quadratic action (\ref{eq:expansion2_SquadMatrix}) is expressed
in the basis of weights and harmonics (\ref{eq:expansion5_basisFields1})
-- (\ref{eq:expansion5_basisFields4}), and the zero modes $\sigma_{0},\tilde{\sigma}_{0},\star F_{0},H_{0}$
are replaced by the moduli $\mathfrak{\mathfrak{m}}$ and $u$ described
in (\ref{eq:locus8_moduliVplet}). As a result, the quadratic action
reduces to 
\begin{equation}
\tilde{S}_{\text{chi}}=\int\text{d}\mu\sum_{\rho,j,j_{3}}\left(\overline{\Phi}^{\rho}_{j,j_{3}}\Delta^{\rho}_{\text{B}}\Phi^{\rho}_{j,j_{3}}+\overline{\Psi}^{\rho}_{j,j_{3}}\Delta^{\rho}_{\text{F}}\Psi^{\rho}_{j,j_{3}}\right),
\end{equation}
where 
\begin{eqnarray}
\Delta^{\rho}_{\mathrm{B}}\Phi^{\rho}_{j,j_{3}} & = & \left(\begin{array}{cccc}
-\frac{1}{2}D^{0}_{\mu}D^{\mu}_{0}+\frac{\rho(\mathfrak{m})}{4R^{2}}-\frac{i\rho(\mathfrak{m})\rho(u)}{4R^{2}t} & 0 & 0 & 0\\
0 & -\frac{1}{2}D^{0}_{\mu}D^{\mu}_{0}-\frac{\rho(\mathfrak{m})}{4R^{2}}+\frac{i\rho(\mathfrak{m})\rho(u)}{4R^{2}t} & 0 & 0\\
0 & 0 & \frac{1}{2} & 0\\
0 & 0 & 0 & \frac{1}{2}
\end{array}\right)\left(\begin{array}{l}
\bar{\phi}^{\rho}_{j,j_{3}}Y^{-\rho(\mathfrak{m})/2}_{j,j_{3}}\\
\phi^{\rho}_{j,j_{3}}Y^{-\rho(\mathfrak{m})/2}_{j,j_{3}}\\
f^{\rho}_{j,j_{3}}Y^{-\rho(\mathfrak{m})/2-1}_{j,j_{3}}\\
\bar{f}^{\rho}_{j,j_{3}}Y^{-\rho(\mathfrak{m})/2+1}_{j,j_{3}}
\end{array}\right),\label{eq:oneLoop_Obos}\\
\Delta^{\rho}_{\mathrm{F}}\Psi^{\rho}_{j,j_{3}} & = & \left(\begin{array}{cccc}
-\frac{\rho(\mathfrak{m})}{R^{2}t} & 0 & 0 & iD^{0}_{\bar{z}}\\
0 & -\frac{\rho(\mathfrak{m})}{R^{2}t} & -iD^{0}_{z} & 0\\
0 & iD^{0}_{\bar{z}} & -\frac{i}{2}\rho(u) & 0\\
-iD^{0}_{z} & 0 & 0 & -\frac{i}{2}\rho(u)
\end{array}\right)\left(\begin{array}{l}
\overline{\psi}^{\rho}_{j,j_{3}}Y^{-\rho(\mathfrak{m})/2}_{j,j_{3}}\\
\psi^{\rho}_{j,j_{3}}Y^{-\rho(\mathfrak{m})/2}_{j,j_{3}}\\
\zeta^{\rho}_{j,j_{3}}Y^{-\rho(\mathfrak{m})/2-1}_{j,j_{3}}\\
\overline{\zeta}^{\rho}_{j,j_{3}}Y^{-\rho(\mathfrak{m})/2+1}_{j,j_{3}}
\end{array}\right).\label{eq:oneLoop_Ofer}
\end{eqnarray}

\subsubsection{Mode analysis\label{subsec:Mode-analysis}}

The procedure to evaluate the integrals over fluctuating modes in
terms of determinants is as follows. First we specify $j$, determine
which of the harmonics exist using (\ref{eq:harmonics}), and remove
rows and columns from the fluctuation operators (\ref{eq:oneLoop_Obos})
and (\ref{eq:oneLoop_Ofer}) if necessary. Following this, we compute
the eigenvalues of $D^{0}_{z}$ and $D^{0}_{\bar{z}}$ using (\ref{eq:eigenvalues}),
then evaluate the determinants and their ratio. 

For $j\geq|\rho(\mathfrak{m})|/2+1,\;\rho(\mathfrak{m})\in\mathbb{Z}$,
the harmonics $Y^{-\rho(\mathfrak{m})/2-1}_{j,j_{3}},\;Y^{-\rho(\mathfrak{m})/2}_{j,j_{3}},\;Y^{-\rho(\mathfrak{m})/2+1}_{j,j_{3}}$
exist, and the eigenvalues have multiplicity $2j+1$. We replace $D^{0}_{z}$
and $D^{0}_{\bar{z}}$ in (\ref{eq:oneLoop_Obos}) and (\ref{eq:oneLoop_Ofer})
with their eigenvalues, evaluate the determinant of each operator,
then their ratio. The ratio of fluctuation determinants for $j\geq|\rho(\mathfrak{m})|/2+1,\;\rho(\mathfrak{m})\in\mathbb{Z}$
is 
\begin{eqnarray}
\left.\left(\frac{\sqrt{\det\Delta^{\rho}_{\text{F}}}}{\sqrt{\det\Delta^{\rho}_{\text{B}}}}\right)\right|_{{j\geq\frac{\left|\rho(\mathfrak{m})\right|}{2}+1\atop \rho(\mathfrak{m})\in\mathbb{Z}}} & = & \prod_{\rho\in\mathcal{R}}\prod_{{j\geq\frac{|\rho(\mathfrak{m})|}{2}+1\atop \rho(\mathfrak{m})\in\mathbb{Z}}}\prod^{j}_{j_{3}=-j}\frac{\sqrt{\frac{\left(j(j+1)-\frac{\rho(\mathfrak{m})}{2}\left(1+\frac{\rho(\mathfrak{m})}{2}-\frac{i\rho(u)}{t}\right)\right)\left(j(j+1)+\frac{\rho(\mathfrak{m})}{2}\left(1-\frac{\rho(\mathfrak{m})}{2}+\frac{i\rho(u)}{t}\right)\right)}{16R^{4}}}}{\sqrt{\frac{\left(j(j+1)-\frac{\rho(\mathfrak{m})}{2}\left(1+\frac{\rho(\mathfrak{m})}{2}-\frac{i\rho(u)}{t}\right)\right)\left(j(j+1)+\frac{\rho(\mathfrak{m})}{2}\left(1-\frac{\rho(\mathfrak{m})}{2}+\frac{i\rho(u)}{t}\right)\right)}{16R^{4}}}}\label{eq:oneLoop_Z1FB}\\
 & = & 1.
\end{eqnarray}
As expected, the bosonic and fermionic determinants cancel, resulting
in a contribution of one. 

For $j=\rho(\mathfrak{m})/2\geq1/2\;\rho(\mathfrak{m})\geq1$, the
harmonics $Y^{-\rho(\mathfrak{m})/2}_{j,j_{3}},\;Y^{-\rho(\mathfrak{m})/2+1}_{j,j_{3}}$
exist, and the eigenvalues have multiplicity $2j+1=\rho(\mathfrak{m})+1$.
As $Y^{-\rho(\mathfrak{m})/2-1}_{j,j_{3}}$ doesn't exist, we remove
the third row and column from (\ref{eq:oneLoop_Obos}) and (\ref{eq:oneLoop_Ofer}).
Evaluating the eigenvalues in the bosonic operator, we have
\begin{eqnarray}
\left.\left(\Delta^{\rho}_{\mathrm{B}}\Phi^{\rho}_{j,j_{3}}\right)\right|_{{j=\frac{\rho(\mathfrak{m})}{2},\atop \rho(\mathfrak{m})\geq1}} & = & \left(\begin{array}{ccc}
-\frac{1}{2}D^{0}_{\mu}D^{\mu}_{0}+\frac{\rho(\mathfrak{m})}{4R^{2}}+\frac{i\rho(\mathfrak{m})\rho(u)}{4R^{2}t} & 0 & 0\\
0 & -\frac{1}{2}D^{0}_{\mu}D^{\mu}_{0}-\frac{\rho(\mathfrak{m})}{4R^{2}}+\frac{i\rho(\mathfrak{m})\rho(u)}{4R^{2}t} & 0\\
0 & 0 & \frac{1}{2}
\end{array}\right)\left(\begin{array}{l}
\bar{\phi}^{\rho}_{j,j_{3}}Y^{-\frac{\rho(\mathfrak{m})}{2}}_{j,j_{3}}\\
\phi^{\rho}_{j,j_{3}}Y^{-\frac{\rho(\mathfrak{m})}{2}}_{j,j_{3}}\\
\bar{f}^{\rho}_{j,j_{3}}Y^{-\frac{\rho(\mathfrak{m})}{2}+1}_{j,j_{3}}
\end{array}\right),\\
 & = & \left(\begin{array}{ccc}
\frac{j(j+1)+\frac{\rho(\mathfrak{m})}{2}\left(-\frac{\rho(\mathfrak{m})}{2}+\frac{i\rho(u)}{t}+1\right)}{2R^{2}} & 0 & 0\\
0 & \frac{j(j+1)-\frac{\rho(\mathfrak{m})}{2}\left(\frac{\rho(\mathfrak{m})}{2}-\frac{i\rho(u)}{t}+1\right)}{2R^{2}} & 0\\
0 & 0 & \frac{1}{2}
\end{array}\right)\left(\begin{array}{l}
\bar{\phi}^{\rho}_{j,j_{3}}Y^{-\frac{\rho(\mathfrak{m})}{2}}_{j,j_{3}}\\
\phi^{\rho}_{j,j_{3}}Y^{-\frac{\rho(\mathfrak{m})}{2}}_{j,j_{3}}\\
\bar{f}^{\rho}_{j,j_{3}}Y^{-\frac{\rho(\mathfrak{m})}{2}+1}_{j,j_{3}}
\end{array}\right),\\
 & = & \left(\begin{array}{ccc}
\frac{\rho(\mathfrak{m})}{2R^{2}}+\frac{i\rho(\mathfrak{m})\rho(u)}{4R^{2}t} & 0 & 0\\
0 & \frac{i\rho(\mathfrak{m})\rho(u)}{4R^{2}t} & 0\\
0 & 0 & \frac{1}{2}
\end{array}\right)\left(\begin{array}{l}
\bar{\phi}^{\rho}_{j,j_{3}}Y^{-\frac{\rho(\mathfrak{m})}{2}}_{j,j_{3}}\\
\phi^{\rho}_{j,j_{3}}Y^{-\frac{\rho(\mathfrak{m})}{2}}_{j,j_{3}}\\
\bar{f}^{\rho}_{j,j_{3}}Y^{-\frac{\rho(\mathfrak{m})}{2}+1}_{j,j_{3}}
\end{array}\right).\label{eq:oneLoop_bosOp2a}
\end{eqnarray}
In the second equality, we use (\ref{eq:eigenvalues}) to replace
the $D^{0}_{\mu}D^{\mu}_{0}$ with its eigenvalue, and in the third
equality, we set $j=\rho(\mathfrak{m})/2$. Notice the purely imaginary
eigenvalue $i\rho(\mathfrak{m})\rho(u)/(4R^{2}t)$ in the middle entry
of the matrix in (\ref{eq:oneLoop_bosOp2a}), which is associated
to the mode coefficient $\phi^{\rho}_{j,j_{3}}$. The presence of
this imaginary eigenvalue indicates that the integral over the fluctuation
$\phi^{\prime}$ is oscillatory at $j=\rho(\mathfrak{m})/2\geq1/2$.
We retain this oscillatory integral and evaluate the remaining integrals
over fluctuations as determinants. Accordingly, the operator (\ref{eq:oneLoop_bosOp2a})
is decomposed as
\begin{equation}
\left.\left(\tilde{\Phi}\Delta_{\text{B}}\Phi\right)\right|_{{j=\frac{\rho(\mathfrak{m})}{2},\atop \rho(\mathfrak{m})\geq1}}=(\tilde{\phi}^{\prime})\left(\begin{array}{c}
\frac{i\rho(\mathfrak{m})\rho(u)}{4R^{2}t}\end{array}\right)(\phi^{\prime})+(\phi^{\prime},F^{\prime}_{\bar{z}})\underset{\Delta^{\prime}_{\text{B}}}{\underbrace{\left(\begin{array}{cc}
\frac{\rho(\mathfrak{m})}{2R^{2}}+\frac{i\rho(\mathfrak{m})\rho(u)}{4R^{2}t} & 0\\
0 & \frac{1}{2}
\end{array}\right)}}\left(\begin{array}{c}
\tilde{\phi}^{\prime}\\
F^{\prime}_{z}
\end{array}\right).\label{eq:oneLoop_bosOp2b}
\end{equation}
The first term on the right hand side of the equality is the part
of the fluctuation operator acting on $\phi^{\prime}$, which contains
the purely imaginary eigenvalue. The final term in (\ref{eq:oneLoop_bosOp2b})
is the part of the fluctuation operator acting on the $F^{\prime}_{z}$
and $\tilde{\phi}^{\prime}$, which we denote $\Delta^{\prime}_{\text{B}}$.
The oscillatory integral associated to the purely imaginary eigenvalue
of the fluctuation $\phi^{\prime}$ is 
\begin{equation}
\left.\left(I(u,\mathfrak{m})\right)\right|_{{j=\frac{\rho(\mathfrak{m})}{2},\atop \rho(\mathfrak{m})\geq1}}=\prod_{\rho\in\mathcal{R}}\int\prod^{\frac{|\rho(\mathfrak{m})|}{2}}_{j_{3}=-\frac{|\rho(\mathfrak{m})|}{2}}\frac{\text{d}\,\phi^{\rho}_{j,j_{3}}\;\text{d}\,\overline{\phi}^{\rho}_{j,j_{3}}}{2\pi i}\exp\left[-(\overline{\phi}^{\rho}_{j,j_{3}})\left(\frac{i\rho(\mathfrak{m})\rho(u)}{4R^{2}t}\right)(\phi^{\rho}_{j,j_{3}})\right].\label{eq:oneloop_bosExp2}
\end{equation}
 The remaining integrals over fluctuations are evaluated in terms
of a determinant, which is
\begin{equation}
\left.\left(\sqrt{\det\Delta^{\prime\rho}_{\text{B}}}\right)\right|_{{j=\frac{\rho(\mathfrak{m})}{2},\atop \rho(\mathfrak{m})\geq1}}=\prod_{\rho\in\mathcal{R}}\left(\frac{\rho(\mathfrak{m})}{4R^{2}}\left(1-\frac{i\rho(u)}{2t}\right)\right)^{\frac{\rho(\mathfrak{m})+1}{2}}.\label{eq:oneLoop_detbosOp2b}
\end{equation}
Having evaluated the bosonic contribution, we turn to the fermionic
contribution. The fermionic fluctuation operator is 
\begin{eqnarray}
\left.\left(\Delta^{\rho}_{\mathrm{F}}\Psi^{\rho}_{j,j_{3}}\right)\right|_{{j=\frac{\rho(\mathfrak{m})}{2},\atop \rho(\mathfrak{m})\geq1}} & = & \left(\begin{array}{ccc}
-\frac{\rho(\mathfrak{m})}{4R^{2}t} & 0 & iD^{0}_{\bar{z}}\\
0 & -\frac{\rho(\mathfrak{m})}{4R^{2}t} & 0\\
-iD^{0}_{z} & 0 & -\frac{i\rho(u)}{2}
\end{array}\right)\left(\begin{array}{l}
\overline{\psi}^{\rho}_{j,j_{3}}Y^{-\frac{\rho(\mathfrak{m})}{2}}_{j,j_{3}}\\
\psi^{\rho}_{j,j_{3}}Y^{-\frac{\rho(\mathfrak{m})}{2}}_{j,j_{3}}\\
\overline{\zeta}^{\rho}_{j,j_{3}}Y^{-\frac{\rho(\mathfrak{m})}{2}+1}_{j,j_{3}}
\end{array}\right).\label{eq:oneLoop_ferOp2a}
\end{eqnarray}
This contribution is evaluated in terms of a standard determinant
\begin{equation}
\left.\left(\sqrt{\det\Delta^{\rho}_{\mathrm{F}}}\right)\right|_{{j=\frac{\rho(\mathfrak{m})}{2},\atop \rho(\mathfrak{m})\geq1}}=\prod_{\rho\in\mathcal{R}}\left(-\frac{\rho(\mathfrak{m})^{2}}{16R^{4}t}\left(1-\frac{i\rho(u)}{2t}\right)\right)^{\frac{\rho(\mathfrak{m})+1}{2}}.\label{eq:oneLoop_detferOp2a}
\end{equation}
Combining (\ref{eq:oneloop_bosExp2}), (\ref{eq:oneLoop_detbosOp2b}),
and (\ref{eq:oneLoop_detferOp2a}), the one-loop contribution is 
\begin{equation}
\left.\left(\frac{\sqrt{\det\Delta^{\rho}_{\text{F}}}}{\sqrt{\det\Delta^{\prime\rho}_{\text{B}}}}I(u,\mathfrak{m})\right)\right|_{{j=\frac{\rho(\mathfrak{m})}{2},\atop \rho(\mathfrak{m})\geq1}}=\prod_{\rho\in\mathcal{R}}\left(-\frac{\rho(\mathfrak{m})}{4R^{2}t}\right)^{\frac{\rho(\mathfrak{m})+1}{2}}\int\prod^{\frac{\rho(\mathfrak{m})}{2}}_{j_{3}=-\frac{\rho(\mathfrak{m})}{2}}\frac{\text{d}\phi^{\rho}_{j,j_{3}}\text{d}\overline{\phi}^{\rho}_{j,j_{3}}}{2\pi i}e^{-\overline{\phi}^{\rho}_{j,j_{3}}\left(\frac{i\rho(\mathfrak{m})\rho(u)}{4R^{2}t}\right)\phi^{\rho}_{j,j_{3}}}.\label{eq:oneloop_Z2FB}
\end{equation}
The above expression is the oscillatory integral over $\phi^{\prime}$
times the ratio of determinants arising from integrating out $\tilde{\phi}^{\prime},F^{\prime}_{z},\tilde{\psi}^{\prime},\psi^{\prime},\psi^{\prime}_{z}$. 

For $j=-\frac{\rho(\mathfrak{m})}{2}\geq\frac{1}{2},\;\rho(\mathfrak{m})\leq-1,$
the harmonics $Y^{-\frac{\rho(\mathfrak{m})}{2}}_{j,j_{3}},\;Y^{-\frac{\rho(\mathfrak{m})}{2}-1}_{j,j_{3}}$
exist, and the eigenvalues have multiplicity $-\rho(\mathfrak{m})+1$.
This case is essentially the same as the previous one. As $Y^{-\frac{\rho(\mathfrak{m})}{2}+1}_{j,j_{3}}$
doesn't exist, we remove the fourth row and column from (\ref{eq:oneLoop_Obos})
and (\ref{eq:oneLoop_Ofer}). Evaluating the eigenvalues of the bosonic
operator, we have 
\begin{equation}
\left.\left(\Delta^{\rho}_{\mathrm{B}}\Phi^{\rho}_{j,j_{3}}\right)\right|_{{j=-\frac{\rho(\mathfrak{m})}{2},\atop \rho(\mathfrak{m})\leq-1}}=\left(\begin{array}{ccc}
\frac{i\rho(\mathfrak{m})\rho(u)}{4R^{2}t} & 0 & 0\\
0 & -\frac{\rho(\mathfrak{m})}{2R^{2}}+\frac{i\rho(\mathfrak{m})\rho(u)}{4R^{2}t} & 0\\
0 & 0 & \frac{1}{2}
\end{array}\right)\left(\begin{array}{l}
\bar{\phi}^{\rho}_{j,j_{3}}Y^{-\frac{\rho(\mathfrak{m})}{2}}_{j,j_{3}}\\
\phi^{\rho}_{j,j_{3}}Y^{-\frac{\rho(\mathfrak{m})}{2}}_{j,j_{3}}\\
f^{\rho}_{j,j_{3}}Y^{-\frac{\rho(\mathfrak{m})}{2}-1}_{j,j_{3}}
\end{array}\right).\label{eq:oneloop_bosOp3a}
\end{equation}
The purely imaginary eigenvalue $i\rho(\mathfrak{m})\rho(u)/(4R^{2}t)$
is now associated to the coefficient $\overline{\phi}^{\rho}_{j,j_{3}}$
in the fluctuating mode $\tilde{\phi}^{\prime}$. The matrix in (\ref{eq:oneloop_bosOp3a})
is decomposed into smaller matrices, one containing the imaginary
eigenvalue, and the other containing the remaining eigenvalues. The
integral associated to the imaginary eigenvalue of fluctuation $\tilde{\phi}^{\prime}$
is 
\begin{equation}
\left.\left(I(u,\mathfrak{m})\right)\right|_{{j=-\frac{\rho(\mathfrak{m})}{2},\atop \rho(\mathfrak{m})\leq-1}}=\prod_{\rho\in\mathcal{R}}\int\prod^{-\frac{\rho(\mathfrak{m})}{2}}_{j_{3}=\frac{\rho(\mathfrak{m})}{2}}\frac{\text{d}\,\phi^{\rho}_{j,j_{3}}\;\text{d}\,\overline{\phi}^{\rho}_{j,j_{3}}}{2\pi i}\exp\left[-(\phi^{\rho}_{j,j_{3}})\left(\frac{i\rho(\mathfrak{m})\rho(u)}{4R^{2}t}\right)(\overline{\phi}^{\rho}_{j,j_{3}})\right].\label{eq:oneLoop_bosExp3}
\end{equation}
Evaluating the integrals over $\phi^{\prime}$ and $F^{\prime}_{\bar{z}}$,
we obtain the determinant 
\begin{equation}
\left.\left(\sqrt{\det\Delta^{\prime\prime\rho}_{\text{B}}}\right)\right|_{{j=-\frac{\rho(\mathfrak{m})}{2},\atop \rho(\mathfrak{m})\leq-1}}=\prod_{\rho\in\mathcal{R}}\left(-\frac{\rho(\mathfrak{m})}{4R^{2}}\left(1-\frac{i\rho(u)}{2t}\right)\right)^{\frac{1-\rho(\mathfrak{m})}{2}},\label{eq:oneLoop_detbosOp3b}
\end{equation}
where $\Delta^{\prime\prime\rho}_{\text{B}}$ is the matrix (\ref{eq:oneloop_bosOp3a})
without the first row and column. The fermionic fluctuation operator
is
\begin{eqnarray}
\left.\left(\Delta^{\rho}_{\mathrm{F}}\Psi^{\rho}_{j,j_{3}}\right)\right|_{{j=-\frac{\rho(\mathfrak{m})}{2},\atop \rho(\mathfrak{m})\leq-1}} & = & \left(\begin{array}{ccc}
-\frac{\rho(\mathfrak{m})}{4R^{2}t} & 0 & 0\\
0 & -\frac{\rho(\mathfrak{m})}{R^{2}t} & -iD^{0}_{z}\\
0 & iD^{0}_{\bar{z}} & -\frac{i}{2}\rho(u)
\end{array}\right)\left(\begin{array}{l}
\overline{\psi}^{\rho}_{j,j_{3}}Y^{-\frac{\rho(\mathfrak{m})}{2}}_{j,j_{3}}\\
\psi^{\rho}_{j,j_{3}}Y^{-\frac{\rho(\mathfrak{m})}{2}}_{j,j_{3}}\\
\zeta^{\rho}_{j,j_{3}}Y^{-\frac{\rho(\mathfrak{m})}{2}-1}_{j,j_{3}}
\end{array}\right),\label{eq:oneLoop_ferOp3a}
\end{eqnarray}
and its determinant is 
\begin{equation}
\left.\left(\sqrt{\det\Delta^{\rho}_{\mathrm{F}}}\right)\right|_{{j=-\frac{\rho(\mathfrak{m})}{2},\atop \rho(\mathfrak{m})\leq-1}}=\prod_{\rho\in\mathcal{R}}\left(-\frac{\rho(\mathfrak{m})^{2}}{16R^{4}t}\left(1-\frac{i\rho(u)}{2t}\right)\right)^{\frac{1-\rho(\mathfrak{m})}{2}}.\label{eq:oneLoop_detferOp3a}
\end{equation}
Combining (\ref{eq:oneLoop_bosExp3}), (\ref{eq:oneLoop_detbosOp3b}),
and (\ref{eq:oneLoop_detferOp3a}), the one-loop contribution is 
\begin{equation}
\left.\left(\frac{\sqrt{\det\Delta^{\rho}_{\text{F}}}}{\sqrt{\det\Delta^{\prime\prime\rho}_{\text{B}}}}I(u,\mathfrak{m})\right)\right|_{{j=-\frac{\rho(\mathfrak{m})}{2},\atop \rho(\mathfrak{m})\leq-1}}=\prod_{\rho\in\mathcal{R}}\left(-\frac{\rho(\mathfrak{m})}{4R^{2}t}\right)^{\frac{1-\rho(\mathfrak{m})}{2}}\int\prod^{-\frac{\rho(\mathfrak{m})}{2}}_{j_{3}=\frac{\rho(\mathfrak{m})}{2}}\frac{\text{d}\phi^{\rho}_{j,j_{3}}\text{d}\overline{\phi}^{\rho}_{j,j_{3}}}{2\pi i}e^{-\phi^{\rho}_{j,j_{3}}\left(\frac{i\rho(\mathfrak{m})\rho(u)}{4R^{2}t}\right)\overline{\phi}^{\rho}_{j,j_{3}}}.\label{eq:oneLoop_Z3FB}
\end{equation}

For $j=\frac{\rho(\mathfrak{m})}{2}-1\geq0,\;\rho(\mathfrak{m})\geq2,$
the harmonic $Y^{-\rho(\mathfrak{m})/2+1}_{j,j_{3}}$ exists, and
the eigenvalues have multiplicity $\rho(\mathfrak{m})-1$. We remove
the first, second and third row and column of (\ref{eq:expansion3_Obos})
and (\ref{eq:expansion4_Ofer}). The bosonic operator is 
\begin{equation}
\left.\left(\Delta^{\rho}_{\mathrm{B}}\Phi^{\rho}_{j,j_{3}}\right)\right|_{{j=\frac{\rho(\mathfrak{m})}{2}-1,\atop \rho(\mathfrak{m})\geq2}}=\left(\frac{1}{2}\right)\left(\bar{f}^{\rho}_{j,j_{3}}Y^{-\frac{\rho(\mathfrak{m})}{2}+1}_{j,j_{3}}\right),
\end{equation}
and its determinant is 
\begin{equation}
\left.\left(\sqrt{\det\Delta^{\rho}_{\mathrm{B}}}\right)\right|_{{j=\frac{\rho(\mathfrak{m})}{2}-1,\atop \rho(\mathfrak{m})\geq2}}=\prod_{\rho\in\mathcal{R}}\left(\frac{1}{2}\right)^{\frac{\rho(\mathfrak{m})-1}{2}}.
\end{equation}
The fermionic operator is
\begin{equation}
\left.\left(\Delta^{\rho}_{\mathrm{F}}\Psi^{\rho}_{j,j_{3}}\right)\right|_{{j=\frac{\rho(\mathfrak{m})}{2}-1,\atop \rho(\mathfrak{m})\geq2}}=\left(-\frac{i\rho(u)}{2}\right)\left(\overline{\zeta}^{\rho}_{j,j_{3}}Y^{-\frac{\rho(\mathfrak{m})}{2}+1}_{j,j_{3}}\right),
\end{equation}
and its determinant is 
\begin{equation}
\left.\left(\sqrt{\det\Delta^{\rho}_{\mathrm{F}}}\right)\right|_{{j=\frac{\rho(\mathfrak{m})}{2}-1,\atop \rho(\mathfrak{m})\geq2}}=\prod_{\rho\in\mathcal{R}}\left(-\frac{i\rho(u)}{2}\right)^{\frac{\rho(\mathfrak{m})-1}{2}}.
\end{equation}
The ratio of fluctuation determinants for $j=\rho(\mathfrak{m})/2-1\geq0$
is 
\begin{equation}
\left.\left(\frac{\sqrt{\det\Delta^{\rho}_{\mathrm{F}}}}{\sqrt{\det\Delta^{\rho}_{\mathrm{B}}}}\right)\right|_{{j=\frac{\rho(\mathfrak{m})}{2}-1,\atop \rho(\mathfrak{m})\geq2}}=\prod_{\rho\in\mathcal{R}}\frac{1}{\left(-i\rho(u)\right)^{\frac{1-\rho(\mathfrak{m})}{2}}}.\label{eq:oneLoop_Z4FB}
\end{equation}

For $j=-\frac{\rho(\mathfrak{m})}{2}-1\geq0,\;\rho(\mathfrak{m})\leq-2,$
the harmonic $Y^{-\rho(\mathfrak{m})/2-1}_{j,j_{3}}$ exists, and
the eigenvalues have multiplicity $-\rho(\mathfrak{m})-1$. We remove
the first, second, and fourth column and row from (\ref{eq:expansion3_Obos})
and (\ref{eq:expansion4_Ofer}). The bosonic operator is 
\begin{equation}
\left.\left(\Delta^{\rho}_{\mathrm{B}}\Phi^{\rho}_{j,j_{3}}\right)\right|_{{j=-\frac{\rho(\mathfrak{m})}{2}-1,\atop \rho(\mathfrak{m})\leq-2}}=\left(\frac{1}{2}\right)\left(f^{\rho}_{j,j_{3}}Y^{-\frac{\rho(\mathfrak{m})}{2}-1}_{j,j_{3}}\right),
\end{equation}
and its determinant is 
\begin{equation}
\left.\left(\sqrt{\det\Delta^{\rho}_{\mathrm{B}}}\right)\right|_{{j=-\frac{\rho(\mathfrak{m})}{2}-1,\atop \rho(\mathfrak{m})\leq-2}}=\prod_{\rho\in\mathcal{R}}\left(\frac{1}{2}\right)^{-\frac{\rho(\mathfrak{m})+1}{2}}.
\end{equation}
The fermionic operator is
\begin{equation}
\left.\left(\Delta^{\rho}_{\mathrm{F}}\Psi^{\rho}_{j,j_{3}}\right)\right|_{{j=-\frac{\rho(\mathfrak{m})}{2}-1,\atop \rho(\mathfrak{m})\leq-2}}=\left(-\frac{i\rho(u)}{2}\right)\left(\zeta^{\rho}_{j,j_{3}}Y^{-\frac{\rho(\mathfrak{m})}{2}-1}_{j,j_{3}}\right),
\end{equation}
and its determinant is 
\begin{equation}
\left.\left(\sqrt{\det\Delta^{\rho}_{\mathrm{F}}}\right)\right|_{{j=-\frac{\rho(\mathfrak{m})}{2}-1,\atop \rho(\mathfrak{m})\leq-2}}=\prod_{\rho\in\mathcal{R}}\left(-\frac{i\rho(u)}{2}\right)^{-\frac{\rho(\mathfrak{m})+1}{2}}.
\end{equation}
The ratio of fluctuation determinants for $j=-\rho(\mathfrak{m})/2-1\geq0$
is 
\begin{equation}
\left.\left(\frac{\sqrt{\det\Delta^{\rho}_{\mathrm{F}}}}{\sqrt{\det\Delta^{\rho}_{\mathrm{B}}}}\right)\right|_{{j=-\frac{\rho(\mathfrak{m})}{2}-1,\atop \rho(\mathfrak{m})\leq-2}}=\prod_{\rho\in\mathcal{R}}\frac{1}{\left(-i\rho(u)\right)^{\frac{1+\rho(\mathfrak{m})}{2}}}.\label{eq:oneLoop_Z5FB}
\end{equation}

For $j=0=\rho(\mathfrak{m})$ the harmonic $Y^{-\rho(\mathfrak{m})/2}_{j,j_{3}}$
exists, and the eigenvalues have multiplicity 1. Since only $Y^{-\rho(\mathfrak{m})/2}_{j,j_{3}}$
exists, we remove third and fourth column and row from (\ref{eq:expansion3_Obos})
and (\ref{eq:expansion4_Ofer}). The bosonic and fermionic operators
reduce to 
\begin{eqnarray}
\left.\left(\Delta^{\rho}_{\mathrm{B}}\Phi^{\rho}_{j,j_{3}}\right)\right|_{{j=0,\atop \rho(\mathfrak{m})=0}} & = & \left.\left(\begin{array}{cc}
-\frac{D^{0}_{\mu}D^{\mu}_{0}}{2}+\frac{\rho(\mathfrak{m})}{4R^{2}}+\frac{i\rho(\mathfrak{m})\rho(u)}{4R^{2}t} & 0\\
0 & -\frac{D^{0}_{\mu}D^{\mu}_{0}}{2}-\frac{\rho(\mathfrak{m})}{4R^{2}}+\frac{i\rho(\mathfrak{m})\rho(u)}{4R^{2}t}
\end{array}\right)\left(\begin{array}{c}
\bar{\phi}^{\rho}_{j,j_{3}}Y^{-\frac{\rho(\mathfrak{m})}{2}}_{j,j_{3}}\\
\phi^{\rho}_{j,j_{3}}Y^{-\frac{\rho(\mathfrak{m})}{2}}_{j,j_{3}}
\end{array}\right)\right|_{{j=0,\atop \rho(\mathfrak{m})=0}},\\
\left.\left(\Delta^{\rho}_{\mathrm{F}}\Psi^{\rho}_{j,j_{3}}\right)\right|_{{j=0,\atop \rho(\mathfrak{m})=0}} & = & \left.\left(\begin{array}{cc}
-\frac{\rho(\mathfrak{m})}{4R^{2}t} & 0\\
0 & -\frac{\rho(\mathfrak{m})}{4R^{2}t}
\end{array}\right)\left(\begin{array}{c}
\overline{\psi}^{\rho}_{j,j_{3}}Y^{-\frac{\rho(\mathfrak{m})}{2}}_{j,j_{3}}\\
\psi^{\rho}_{j,j_{3}}Y^{-\frac{\rho(\mathfrak{m})}{2}}_{j,j_{3}}
\end{array}\right)\right|_{{j=0,\atop \rho(\mathfrak{m})=0}},
\end{eqnarray}
respectively, where all the entries of both of the matrices evaluate
to zero. The primary issue here is that the ratio of determinants
$\sqrt{\det\Delta^{\rho}_{\text{F}}}/\sqrt{\det\Delta^{\rho}_{\text{B}}}$
involves a zero in the denominator. On the one hand, we cannot rigorously
evaluate the contribution at $j=\rho(\mathfrak{m})=0$ due to the
singular bosonic contribution. On the other hand, we can speculate
on the value of this contribution by considering fixed $\rho(\mathfrak{m})/t$
as $t\to0$. In this case, the zero-divided-by-zero cancellation in
the ratio of determinants becomes a zero-divided-by-zero cancellation
in the $t$-dependent terms. Keeping $\rho(\mathfrak{m})/t$ fixed
while setting all other $j$ and $\rho(\mathfrak{m})$ to zero, the
bosonic and fermionic operators are 
\begin{eqnarray}
\left.\left(\hat{\Delta}^{\rho}_{\mathrm{B}}\Phi^{\rho}_{j,j_{3}}\right)\right|_{{j=0,\atop \rho(\mathfrak{m})=0}} & = & \left(\begin{array}{cc}
\frac{i\rho(\mathfrak{m})\rho(u)}{4R^{2}t} & 0\\
0 & \frac{i\rho(\mathfrak{m})\rho(u)}{4R^{2}t}
\end{array}\right)\left(\begin{array}{c}
\bar{\phi}^{\rho}_{0,0}Y^{0}_{0,0}\\
\phi^{\rho}_{0,0}Y^{0}_{0,0}
\end{array}\right),
\end{eqnarray}
\begin{equation}
\left.\left(\hat{\Delta}^{\rho}_{\mathrm{F}}\Psi^{\rho}_{j,j_{3}}\right)\right|_{{j=0,\atop \rho(\mathfrak{m})=0}}=\left(\begin{array}{cc}
-\frac{\rho(\mathfrak{m})}{4R^{2}t} & 0\\
0 & -\frac{\rho(\mathfrak{m})}{4R^{2}t}
\end{array}\right)\left(\begin{array}{c}
\overline{\psi}^{\rho}_{0,0}Y^{0}_{0,0}\\
\psi^{\rho}_{0,0}Y^{0}_{0,0}
\end{array}\right),
\end{equation}
and the ratio of determinants is 
\begin{equation}
\left.\left(\frac{\sqrt{\det\hat{\Delta}^{\rho}_{\mathrm{F}}}}{\sqrt{\det\hat{\Delta}^{\rho}_{\mathrm{B}}}}\right)\right|_{{j=0,\atop \rho(\mathfrak{m})=0}}=\begin{cases}
\;\prod_{\rho\in\mathcal{R}}\frac{1}{\sqrt{\left(-i\rho(u)\right)^{2}}}, & \text{as }t\to0,\\
\;\text{singular}, & \text{for }t>0.
\end{cases}\label{eq:Z1L_zeroFlux}
\end{equation}
In particular, the factors of $\rho(\mathfrak{m})/(4R^{2}t)$ cancel
and we are left with a contribution that resembles those of other
sectors. Despite the similarity, setting $t=0$ is problematic in
several ways. For instance, at $t=0$, the $Q_{A}$-exact deformation
in the vector multiplet localizing term (\ref{eq:locTerm2_QVvec})
is turned off, the Yang-Mills equations (\ref{eq:locus6_YMeq}) are
singular, and the Yang-Mills connections are at infinity in field
space. For these reasons, we omit the $j=\rho(\mathfrak{m})=0$ contribution. 

Collecting the results in (\ref{eq:oneLoop_Z1FB}), (\ref{eq:oneloop_Z2FB}),
(\ref{eq:oneLoop_Z3FB}), (\ref{eq:oneLoop_Z4FB}), and (\ref{eq:oneLoop_Z5FB}),
the one-loop contribution is 
\begin{equation}
Z^{\text{chiral}}_{\text{1-loop}}=\prod_{\rho\in\mathcal{R}}\frac{\left(-\frac{\rho(\mathfrak{m})}{4R^{2}t}\right)^{\frac{\rho(\mathfrak{m})+1}{2}}}{\left(-\frac{i\rho(u)}{2}\right)^{\frac{\rho(\mathfrak{m})+1}{2}}}\int\prod^{\frac{\rho(\mathfrak{m})+1}{2}}_{\ell=1}\frac{\text{d}\,\phi^{\rho}_{\ell}\;\text{d}\,\overline{\phi}^{\rho}_{\ell}}{2\pi i}\exp\left[-\overline{\phi}^{\rho}_{\ell}\,\left(\frac{i\rho(\mathfrak{m})\rho(u)}{4R^{2}t}\right)\,\phi^{\rho}_{\ell}\right],\quad\rho(\mathfrak{m})\neq0.\label{eq:Z1Lcorrect}
\end{equation}
The one-loop contribution is a product of a ratio of fluctuation determinants
and an oscillatory integral. The oscillatory integral is associated
to the fluctuating modes of the bosonic scalars at $j=|\rho(\mathfrak{m})|/2,\;|\rho(\mathfrak{m})|\geq1$,
and the ratio of determinants is the result of evaluating all remaining
integrals. 

Finally, let us remark on an erroneous treatment of the one-loop contribution.
Had all the eigenvalues been included in the ratio of fluctuation
determinants, the one-loop contribution would have been 

\begin{equation}
\prod_{\rho\in\mathcal{R}}\frac{1}{\left(-i\rho(u)\right)^{\rho(\mathfrak{m})+1}},\quad\rho(\mathfrak{m})\neq0.
\end{equation}
This result is incorrect. It treats the purely imaginary eigenvalue
$i\rho(\mathfrak{m})\rho(u)/(4R^{2}t)$ as though it were real, and
treats the associated oscillatory integral as though it were Gaussian. 

\subsubsection{The distributional integral\label{subsec:The-distributional-integral} }

In this section, the oscillatory integral in (\ref{eq:Z1Lcorrect})
is reduced to a linear combination of distributions, for the case
of abelian $A$-twisted $\mathcal{N}=(2,2)$ GLSMs with $N$ chiral
multiplets of gauge charge 1 and R-charge 0. This proceeds by first
specializing to gauge group $\text{U}(1)$, then rewriting the complex
integration variables in terms of a real radial integration variable,
and finally using a distributional identity to express the oscillatory
integral as a distribution.

The starting point is the oscillatory integral
\begin{equation}
I(u,\mathfrak{m})=\int\prod_{\rho\in\mathcal{R}}\prod^{\frac{|\rho(\mathfrak{m})|+1}{2}}_{\ell=1}\frac{\text{d}\phi_{\rho,\ell}\,\text{d}\tilde{\phi}_{\rho,\ell}}{2\pi i}\,\exp\left[\mp\tilde{\phi}_{\rho,\ell}\left(i\,\tau\,\rho(\mathfrak{m})\,\rho(u)\right)\phi_{\rho,\ell}\right],\label{eq:singIntNonAbelian}
\end{equation}
where $\rho(\mathfrak{m})\in\mathbb{Z}\backslash\{0\}$, $\rho(u)\in\mathbb{R}$,
and $\tau=1/(4R^{2}t)\in\mathbb{R}_{>0}$. For a $G=\text{U}(1)$
theory with chiral multiplets $\Phi_{i},\,i=1,\dots,N$ of gauge charge
$Q_{i}$, the integral (\ref{eq:singIntNonAbelian}) is 

\begin{equation}
I=\int\prod^{N}_{i=1}\prod^{\frac{\left|Q_{i}\mathfrak{m}\right|+1}{2}}_{\ell=1}\frac{\text{d}\phi_{i,\ell}\,\text{d}\tilde{\phi}_{i,\ell}}{2\pi i}\,\exp\left[\mp\tilde{\phi}_{i,\ell}\left(i\,\tau\,Q_{i}\mathfrak{m}\,Q_{i}u\right)\phi_{i,\ell}\right].\label{eq:singIntAbelian}
\end{equation}
Specializing to $Q_{i}=1$ for $i=1,\dots,N$, the expression becomes
\begin{equation}
I=\int\prod^{N}_{i=1}\prod^{\frac{|\mathfrak{m}|+1}{2}}_{\ell=1}\frac{\text{d}\phi_{i,\ell}\,\text{d}\tilde{\phi}_{i,\ell}}{2\pi i}\,\exp\left[\mp\tilde{\phi}_{i,\ell}\left(i\,\tau\,\mathfrak{m}\,u\right)\phi_{i,\ell}\right].\label{eq:singIntAbelianCharge1}
\end{equation}
We make a change of variables to real coordinates
\begin{eqnarray}
\phi_{i,\ell} & = & -i(\phi_{1\,i,\ell}+i\phi_{2\,i,\ell}),\\
\tilde{\phi}_{i,\ell} & = & i(\phi_{1\,i,\ell}-i\phi_{2\,i,\ell}).
\end{eqnarray}
The measure changes as $\text{d}\phi_{i,\ell}\,\text{d}\tilde{\phi}_{i,\ell}=i\text{d}\phi_{1\,i,\ell}\,\text{d}\phi_{2\,i,\ell}$,
and there are $N$ values of $i$ and $(|\mathfrak{m}|+1)/2$ values
of $\ell$. Consequently, the integral becomes 
\begin{eqnarray}
I & = & \int_{\mathbb{R}}\prod^{\frac{N(|\mathfrak{m}|+1)}{2}}_{\ell=1}\frac{\text{d}\phi_{1\,\ell}\,\text{d}\phi_{2\,\ell}}{2\pi}\,\exp\left[\mp i\,\tau\,\mathfrak{m}\,u\left(\phi^{2}_{1\,\ell}+\phi^{2}_{2\,\ell}\right)\right].\label{eq:singIntAbelianCharge1Real}
\end{eqnarray}
Scaling by $\phi_{\ell}\to\sqrt{2\pi}\phi_{\ell}$, and combining
the integration variables, the integral is 
\begin{equation}
I=\int_{\mathbb{R}}\prod^{K}_{k=1}\text{d}\phi_{k}\,e^{\mp2\pi i\tau\mathfrak{m}u\phi^{2}_{k}},\label{eq:singIntAbelianCharge1RealScaled}
\end{equation}
where $K=N(|\mathfrak{m}|+1)$. We introduce radial coordinates 
\begin{equation}
r^{2}=\sum^{K}_{k=1}\phi^{2}_{k},\qquad\prod^{K}_{k=1}\text{d}\phi_{k}=r^{K-1}\text{d}r\,\text{d}\Omega_{K-1},\label{eq:singIntRadialr}
\end{equation}
where $\text{d}\Omega_{K-1}$ is the surface area element of the $K-1$
dimensional unit sphere. Expressing (\ref{eq:singIntAbelianCharge1RealScaled})
in terms of (\ref{eq:singIntRadialr}), we have 
\begin{equation}
I=\text{vol}\Omega_{K-1}\int^{\infty}_{0}\text{d}r\,r^{K-1}e^{\mp2\pi i\tau\mathfrak{m}ur^{2}}.\label{eq:singIntAbelianCharge1RealScaledRadialr}
\end{equation}
where $\text{vol}\Omega_{K-1}=2\pi^{K/2}/\Gamma(K/2)$ is the volume
of the $K-1$ dimensional unit sphere. Introducing a second radial
coordinate, 
\begin{equation}
y=r^{2},\qquad\text{d}r=\frac{\text{d}y}{2\sqrt{y}},\label{eq:singIntradialy}
\end{equation}
the integral (\ref{eq:singIntAbelianCharge1RealScaledRadialr}) becomes
\begin{eqnarray}
I & = & \frac{\pi^{\frac{K}{2}}}{\Gamma\left(\frac{K}{2}\right)}\int^{\infty}_{0}\text{d}y\,y^{\frac{K}{2}-1}e^{\mp2\pi i\tau\mathfrak{m}uy}.\label{singIntAbelianCharge1RealScaledRadialy}
\end{eqnarray}
For $K\geq2$, the integrand may be expressed as 
\begin{equation}
y^{\frac{K}{2}-1}e^{-2\pi i\tau\mathfrak{m}uy}=\left(\frac{1}{\mp2\pi i\tau\mathfrak{m}}\frac{\text{d}}{\text{d}u}\right)^{\frac{K}{2}-1}e^{\mp2\pi i\tau\mathfrak{m}uy},
\end{equation}
such that the integral (\ref{singIntAbelianCharge1RealScaledRadialy})
reads
\begin{eqnarray}
I & = & \frac{\pi^{\frac{K}{2}}}{\Gamma\left(\frac{K}{2}\right)}\left(\frac{1}{\mp2\pi i\,\tau\,\mathfrak{m}}\frac{\text{d}}{\text{d}u}\right)^{\frac{K}{2}-1}\int^{\infty}_{0}\text{d}ye^{\mp2\pi i\tau\mathfrak{m}uy}.\label{singIntAbelianCharge1RealScaledRadialyDer}
\end{eqnarray}
Now, using the Fourier transform of the Heaviside step function in
the sense of tempered distributions\footnote{The Fourier transform is 
\[
\hat{\Theta}(p)=\lim_{b\to\infty}\int^{b}_{-b}e^{\mp2\pi ixp}\Theta(x)\,\text{d}x=\frac{1}{2}\left(\delta(p)\mp\frac{i}{\pi}\text{pv}\frac{1}{p}\right),\qquad \Theta(x)=\begin{cases}
1, & x\geq0,\\
0, & x<0.
\end{cases}
\]
}, the final integral is
\begin{eqnarray}
\int^{\infty}_{0}\text{d}ye^{\mp2\pi i\tau\mathfrak{m}uy} & = & \frac{1}{\pm2\pi i}\text{pv}\left(\frac{1}{\tau\,\mathfrak{m}\,u}\right)+\frac{\delta(\tau\,\mathfrak{m}\,u)}{2}\\
 & = & \frac{1}{\pm2\pi i\,\tau\,\mathfrak{m}}\text{pv}\frac{1}{u}+\frac{\delta(u)}{2\left|\tau\,\mathfrak{m}\right|}\\
 & = & \frac{1}{\pm2\pi i\,\tau\,\mathfrak{m}}\left(\text{pv}\frac{1}{u}\pm i\pi\,\text{sgn}(\mathfrak{m})\,\delta(u)\right).
\end{eqnarray}
Consequently, (\ref{singIntAbelianCharge1RealScaledRadialyDer}) reduces
to 
\begin{eqnarray}
I & = & -\frac{1}{\Gamma\left(\frac{K}{2}\right)\left(\mp2i\,\tau\,\mathfrak{m}\right)^{\frac{K}{2}}}\left(\frac{\text{d}}{\text{d}u}\right)^{\frac{K}{2}-1}\left(\text{pv}\frac{1}{u}\pm i\pi\,\text{sgn}(\mathfrak{m})\,\delta(u)\right),\label{eq:singIntAbelianCharge1RealScaledPV0}
\end{eqnarray}
Collecting the results, we found three equivalent expressions for
the oscillatory integral (\ref{eq:singIntAbelianCharge1}) in the
abelian case. These are
\begin{eqnarray}
I(u,\mathfrak{m}) & = & \int\prod^{K}_{k=1}\frac{\text{d}\phi_{k}}{\sqrt{2\pi}}\,e^{\mp\,i\,\tau\,\mathfrak{m}\,u\,\phi^{2}_{k}},\label{eq:CPN_Ireal}\\
 & = & \frac{\text{vol}\Omega_{K-1}}{2}\int^{\infty}_{0}\text{d}y\,y^{\frac{K}{2}-1}e^{\mp\,2\pi i\,\tau\,\mathfrak{m}\,u\,y},\label{eq:CPN_Iradial}\\
 & = & -\frac{1}{\Gamma\left(\frac{K}{2}\right)\left(\mp2i\,\tau\,\mathfrak{m}\right)^{\frac{K}{2}}}\frac{\text{d}^{\frac{K}{2}-1}}{\text{d}u^{\frac{K}{2}-1}}\left(\text{pv}\frac{1}{u}\pm i\pi\,\text{sgn}(\mathfrak{m})\,\delta(u)\right),
\end{eqnarray}
where the final expression is a linear combination of tempered distributions.
Here, $K=N(|\mathfrak{m}|+1)$ and $\tau=1/(4R^{2}t)$ with $\mathfrak{m\in\mathbb{Z}}\backslash\{0\}$,
$N\in\mathbb{Z}_{>1}$ and $t\in\mathbb{R}_{>0}$. 

Finally, let us note that the integral (\ref{eq:CPN_Ireal}) appears
in equation 2.10 of \cite{Witten:1988hf}, in the context of Chern-Simons
theory. There, Witten expressed the integral in terms of the $\eta$-invariant
of Atiyah, Patodi, and Singer \cite{Atiyah:1975jf}, then related
it to the phase of the path integral of Chern-Simons theory. It would
be interesting to understand whether the $\eta$-invariant perspective
offers any insight or utility. 

\subsection{Results of localization \label{subsec:Results-of-localization}}

Observables of $A$-twisted $\mathcal{N}=(2,2)$ theories of vector
and chiral multiplets on $S^{2}$ with gauge group $G$ are described
by the formula 
\begin{equation}
\left\langle \mathcal{O}\right\rangle =\frac{1}{|W|}\,\sum_{\mathfrak{m}\in\Lambda^{G}_{\text{cochar}}}\,\int_{\mathbb{R}}\text{d}u^{r}\,\mathcal{O}(u)\,Z_{\text{cl}}(u,\mathfrak{m})\,Z_{\text{1-loop}}(u,\mathfrak{m}).\label{eq:oneLoop_correlatorGeneric}
\end{equation}
Here, $\mathfrak{m}\in\Lambda^{G}_{\text{cochar}}\subset\mathfrak{h}$
is the GNO quantized gauge flux on $S^{2}$, $\Lambda^{G}_{\text{cochar}}$
is the cocharacter lattice of $G$, $\mathfrak{h}$ is the Cartan
subalgebra of $\mathfrak{g}=\text{Lie}\,G$, $u\in\mathfrak{h}$ parameterizes
the bosonic scalar $\sigma$ in the vector multiplet, $\mathcal{O}$
is a $Q_{A}$-closed gauge invariant operator insertion, $Z_{\text{cl}}$
is the classical contribution, $Z_{\text{1-loop}}$ is the one-loop
contribution, $|W|$ is the order of the Weyl group of $G$, and $r=\text{rank}\,G$.

The classical contribution from the $Q_{A}$-closed F-term action
of the vector multiplet is
\begin{equation}
Z_{\text{cl}}=e^{4\pi\,\text{Tr}\,\widetilde{W}^{\prime}(u)\,\mathfrak{m}},
\end{equation}
where the twisted superpotential $\widetilde{W}(u)$ is holomorphic,
and the prime denotes differentiation \cite{Leeb-Lundberg:2023jsj}.
Upon choosing linear $\widetilde{W}(u)$, such that it represents
a complexified Fayet-Iliopoulos term, the classical contribution becomes
\begin{equation}
Z^{(\text{FI})}_{\text{cl}}=\xi^{\text{Tr}\,\mathfrak{m}},\qquad\xi=e^{-(\zeta+i\theta)}.
\end{equation}
The one-loop contributions of vector multiplets is 
\begin{equation}
Z^{\text{vector}}_{\text{1-loop}}=(-1)^{\sum_{\alpha>0}\alpha(\mathfrak{m})}\prod_{\alpha}\alpha(u),
\end{equation}
where $\alpha$ are roots (weights of the adjoint representation)
\cite{Leeb-Lundberg:2023jsj}. The one-loop contribution of chiral
multiplets is 
\begin{equation}
Z^{\text{chiral}}_{\text{1-loop}}=\prod_{\rho\in\mathcal{R}}\frac{\left(-\frac{\rho(\mathfrak{m})}{4R^{2}t}\right)^{\frac{\rho(\mathfrak{m})+1}{2}}}{\left(-\frac{i\rho(u)}{2}\right)^{\frac{\rho(\mathfrak{m})+1}{2}}}\underset{I(u,\mathfrak{m})}{\underbrace{\int\prod^{\frac{\rho(\mathfrak{m})+1}{2}}_{\ell=1}\frac{\text{d}\,\phi^{\rho}_{\ell}\;\text{d}\,\overline{\phi}^{\rho}_{\ell}}{2\pi i}\exp\left[-\overline{\phi}^{\rho}_{\ell}\,\left(\frac{i\rho(\mathfrak{m})\rho(u)}{4R^{2}t}\right)\,\phi^{\rho}_{\ell}\right]}},\quad\rho(\mathfrak{m})\neq0,\label{eq:oneLoopResults_Z1Ldistr0}
\end{equation}
where the oscillatory integral $I(u,\mathfrak{m})$ is associated
to the fluctuating modes of the bosonic scalars at $j=|\rho(\mathfrak{m})|/2$,
$|\rho(\mathfrak{m})|\geq1$. For $G=U(1)$ theories with $N$ chiral
multiplets of gauge charge 1 and R-charge 0, (\ref{eq:oneLoopResults_Z1Ldistr0})
reduces to 
\begin{eqnarray}
Z^{\text{\text{chiral}}}_{\text{1-loop}} & = & \frac{\left(-\frac{\mathfrak{m}}{2R^{2}t}\right)^{\frac{N(\mathfrak{m}+1)}{2}}}{\left(-iu\right)^{\frac{N(\mathfrak{m}+1)}{2}}}\cdot I(u,\mathfrak{m}),\label{eq:eq:oneLoopResults_Z1Ldistr1}
\end{eqnarray}
where 
\begin{equation}
I(u,\mathfrak{m})=-\frac{1}{\left(-\frac{i\mathfrak{m}}{2R^{2}t}\right)^{\frac{N(\mathfrak{m}+1)}{2}}\Gamma\left(\frac{N(\mathfrak{m}+1)}{2}\right)}\frac{\text{d}^{\frac{N(\mathfrak{m}+1)}{2}-1}}{\text{d}u^{\frac{N(\mathfrak{m}+1)}{2}-1}}\left(\text{pv}\frac{1}{u}+i\pi\text{sgn}(\mathfrak{m})\delta(u)\right),\label{eq:eq:oneLoopResults_Z1Ldistr2}
\end{equation}
for $N(\mathfrak{m}+1)\geq2$ and $\mathfrak{m}\in\mathbb{Z}\backslash\{0\}$.
Note that (\ref{eq:eq:oneLoopResults_Z1Ldistr2}) may be equivalently
expressed in terms of either a real integral (\ref{eq:CPN_Ireal}),
or a radial integral (\ref{eq:CPN_Iradial}). 

\section{The $A$-twisted $\mathbb{CP}^{N-1}$ model \label{sec:CPN}}

In this section, we verify the distributional integral description
of observables by computing the correlator of the $A$-twisted $\mathbb{CP}^{N-1}$
model, confirming the standard selection rule, and then recovering
the established complex contour integral description via hyperfunctions.
This proceeds by first reviewing the established complex contour integral
description, presenting the distributional integral description, then
evaluating the distributional integral. Following this, we interpret
the distributional correlator in terms of hyperfunctions, recover
the complex contour integral description, and find agreement with
the JK residue prescription. 

\subsection{Complex contour integral description (review) \label{subsec:ComplexDescription}}

The $A$-twisted $\mathbb{CP}^{N-1}$ model is an abelian $A$-twisted
$\mathcal{N}=(2,2)$ GLSM on $S^{2}$ with $N$ chiral multiplets
of gauge charge $1$ and vector-like R-charge $0$. The GLSM flows
in the IR to an NLSM with target $\mathbb{CP}^{N-1}$. The observables
of interest are correlators of the field-strength twisted chiral multiplet
$\Sigma$, whose lowest component, $\sigma$, is parameterized by
the continuous modulus $u$. Insertions of $\Sigma(x)$ at points
$x$ on $S^{2}$ represent the Kähler class of $\mathbb{CP}^{N-1}$.
The correlator of the $A$-twisted $\mathbb{CP}^{N-1}$ model is 

\begin{equation}
\left\langle \prod^{s}_{a=1}\Sigma_{a}(x_{a})\right\rangle =\sum_{\mathfrak{m}\in\mathbb{Z}}\int_{\mathcal{C}_{\text{JK}}}\frac{\text{d}u}{2\pi i}\frac{u^{s}\xi^{\mathfrak{m}}}{u^{N(\mathfrak{m}+1)}}=\begin{cases}
\xi^{\frac{s-N+1}{N}} & s=N-1\quad(\text{mod }N)\\
0 & \text{otherwise}
\end{cases}.\label{eq:CPN_correlatorJK1}
\end{equation}
Here, $\mathfrak{m}$ is the quantized $\text{U}(1)$ gauge flux on
$S^{2}$, $u$ parameterizes the bosonic scalar $\sigma$ in the vector
multiplet, $\mathcal{C}_{\text{JK}}$ is the complex contour specified
by the Jeffrey-Kirwan (JK) residue prescription, $s$ is the number
of insertions, the complexified Fayet-Iliopoulos (FI) term $\xi^{\mathfrak{m}}=e^{-(\zeta+i\theta)\mathfrak{m}}$
is the classical contribution, and $u^{N(\mathfrak{m}+1)}$ is the
one-loop contribution without twisted masses. For references, see
page 54 of \cite{Benini:2015noa}, page 72 of \cite{Closset:2015rna},
and page 25 of \cite{Benini:2016hjo}.

The evaluation of the integral in (\ref{eq:CPN_correlatorJK1}) requires
a choice of the JK parameter $\eta$. For $\eta>0$, the contour $\mathcal{C}_{\text{JK}}$
encloses the pole at $u=0$. The corresponding residue is then equal
to one when the selection rule 

\begin{equation}
s=N(\mathfrak{m}+1)-1\label{eq:CPN_selectionRule}
\end{equation}
is satisfied, and vanishes otherwise. In what follows, we verify our
localization computation by first reproducing the selection rule (\ref{eq:CPN_selectionRule}),
then recovering a complex contour integral description of the correlator
that coincides with the one in (\ref{eq:CPN_correlatorJK1}). 

\subsection{Distributional integral description \label{subsec:DistributionalDescription}}

In this section, we evaluate the correlator of the $A$-twisted $\mathbb{CP}^{N-1}$
model (\ref{eq:CPN_correlatorJK1}) in the framework of distribution
theory and reproduce the selection rule (\ref{eq:CPN_selectionRule}).
We specialize the formula for observables (\ref{eq:oneLoop_correlatorGeneric})
to the case of the $\mathbb{CP}^{N-1}$ correlator by taking gauge
group $G=U(1)$, operator insertion $\mathcal{O}=\prod^{s}_{a=1}\Sigma_{a}(x_{a})$,
classical contribution $Z_{\text{cl}}=\xi^{\mathfrak{m}}$, and one-loop
contribution (\ref{eq:eq:oneLoopResults_Z1Ldistr1}). Expressing the
correlator as a sum over fluxes,
\begin{eqnarray}
\left\langle \prod^{s}_{a=1}\Sigma_{a}(x_{a})\right\rangle  & = & \sum_{\mathfrak{m}\in\mathbb{Z}}\left\langle \prod^{s}_{a=1}\Sigma_{a}(x_{a})\right\rangle _{(\mathfrak{m})},\label{eq:CPN_correlatorDist}
\end{eqnarray}
the analysis of section \ref{sec:Localization} determines 
\begin{equation}
\left\langle \prod^{s}_{a=1}\Sigma_{a}(x_{a})\right\rangle _{(\mathfrak{m})}=\frac{\xi^{\mathfrak{m}}\left(-1\right)^{\frac{N(\mathfrak{m}+1)}{2}-1}}{\Gamma\left(\frac{N(\mathfrak{m}+1)}{2}\right)}\int_{\mathbb{R}}\text{d}u\left(u^{s-\frac{N(\mathfrak{m}+1)}{2}}\frac{\text{d}^{\frac{N}{2}(\mathfrak{m}+1)-1}}{\text{d}u^{\frac{N}{2}(\mathfrak{m}+1)-1}}\left(\text{pv}\frac{1}{u}+i\pi\,\text{sgn}(\mathfrak{m})\,\delta(u)\right)\right),
\end{equation}
for fixed $\mathfrak{m}\in\mathbb{Z}\backslash\{0\}$, $s\in\mathbb{Z}_{>0}$,
and $N\in\mathbb{Z}_{>1}$. For clarity, the integral is denoted
\begin{equation}
\mathcal{I}_{\mathfrak{m}}=\frac{\left(-1\right)^{q}}{q!}\int_{\mathbb{R}}\text{d}u\left(u^{p}\partial^{q}\left(\text{pv}\frac{1}{u}+i\pi\,\text{sgn}(\mathfrak{m})\,\delta(u)\right)\right)\label{eq:CPN_intDistribution}
\end{equation}
where 
\begin{equation}
p=s-\frac{N}{2}(\mathfrak{m}+1),\quad q=\frac{N}{2}(\mathfrak{m}+1)-1,\quad\partial^{q}=\frac{\text{d}^{q}}{\text{d}u^{q}}.\label{eq:CPN_pq}
\end{equation}
To interpret (\ref{eq:CPN_intDistribution}) as a distributional pairing
$\left\langle T,\varphi\right\rangle $, we must specify a test function
$\varphi$. The function $u^{p}$ lacks compact support and rapid
decay when $p\geq0$, and it is singular at $u=0$ when $p<0$. As
the support of $\delta(u)$ is $\{0\}$, the pairing $\left\langle \delta(u),u^{p}\right\rangle $
is well defined for $p\geq0$. In contrast, the support of $\text{pv}\left(1/u\right)$
is $\mathbb{R}$, and its pairing with $u^{p}$ is divergent. To address
this issue, we introduce a Gaussian regulator and consider (\ref{eq:CPN_intDistribution})
to be defined by the $a\to0^{+}$ limit of 
\begin{equation}
\mathcal{I}_{\mathfrak{m},a}=\frac{\left(-1\right)^{q}}{q!}\int_{\mathbb{R}}\text{d}u\left(e^{-au^{2}}u^{p}\partial^{q}\left(\text{pv}\left(\frac{1}{u}\right)+i\pi\,\text{sgn}(\mathfrak{m})\,\delta(u)\right)\right).\label{eq:CPN_intDistributionReg}
\end{equation}
For $a>0$ and $p\geq0$, the Gaussian regulator plays the role of
a test function and renders the integral meaningful in the sense of
tempered distributions. Specifically, (\ref{eq:CPN_intDistributionReg})
has the form $\left\langle T,\varphi\right\rangle $, with $\varphi\left(u\right)=e^{-au^{2}}\in\mathcal{S}$
and $T\in\mathcal{S}^{\prime}$, where $\mathcal{S}$ denotes Schwartz
space and $\mathcal{S}^{\prime}$ denotes the dual space of tempered
distributions. 

The integral (\ref{eq:CPN_intDistributionReg}) admits algebraic singularities
of order $\kappa=q-p+1$ at $u=0$. Since the Cauchy principal value
does not exist for all $\kappa\in\mathbb{Z}_{>0}$, whereas the Hadamard
finite part does, the integral is evaluated using the finite-part
distribution $\text{pf}(1/u)$ \cite{kanwal2012generalized} rather
than the principal-value distribution $\text{pv}(1/u)$. Note that
the Hadamard finite part of an integral coincides with its principal
value, when the principal value exists. Our analysis is restricted
to $p,q\in\mathbb{Z}_{\geq0}$, or equivalently, the subset of fluxes
\begin{equation}
\mathsf{M}=\left\{ \;\mathfrak{m}\in\mathbb{Z}\backslash\{0\}\;\vert\;N(\mathfrak{m}+1)\in2\mathbb{Z},\;s\geq N(\mathfrak{m}+1)/2\geq1\;\right\}.\label{eq:CPN_subset}
\end{equation}
The remaining cases of $\mathfrak{m}\in\mathbb{Z}\backslash\mathsf{M}$
lie beyond the scope of this work, though some could potentially be
addressed via primitives and fractional differentiation \cite{duistermaat2010distributions}.
In what follows, we evaluate the $\delta(u)$ and $1/u$ contributions
to (\ref{eq:CPN_intDistributionReg}) separately, then combine the
results. 

We begin by evaluating the $\delta(u)$ contribution to (\ref{eq:CPN_intDistributionReg}),
which is 
\begin{eqnarray}
\mathcal{I}^{(\delta)}_{\mathfrak{m},a} & = & \frac{i\pi\,\text{sgn}(\mathfrak{m})\,\left(-1\right)^{q}}{q!}\int_{\mathbb{R}}\text{d}u\left(e^{-au^{2}}u^{p}\partial^{q}\delta(u)\right),\qquad p,q\in\mathbb{Z}_{\geq0}.\label{eq:CPN_intDistributionDelta0}
\end{eqnarray}
Applying the distributional identities \cite{kanwal2012generalized}
\begin{equation}
u^{p}\partial^{q}\delta(u)=\begin{cases}
\frac{\left(-1\right)^{p}q!}{\left(q-p\right)!}\delta^{(q-p)}(u), & q\geq p,\\
0, & q<p,
\end{cases}
\end{equation}
to the integral in (\ref{eq:CPN_intDistributionDelta0}), we obtain
\begin{equation}
\int_{\mathbb{R}}\text{d}u\left(e^{-au^{2}}u^{p}\partial^{q}\delta(u)\right)=\begin{cases}
\frac{\left(-1\right)^{p}q!}{\left(q-p\right)!}\int_{\mathbb{R}}\text{d}u\left(e^{-au^{2}}\delta^{(q-p)}(u)\right), & q\geq p,\\
0, & q<p.
\end{cases}\label{eq:CPN_intDistributionDelta1}
\end{equation}
The integral on the right hand side of the equality reduces to 
\begin{equation}
\int_{\mathbb{R}}\text{d}u\left(e^{-au^{2}}\delta^{(q-p)}(u)\right)=\left(-1\right)^{q-p}\left.\partial^{q-p}\left(e^{-au^{2}}\right)\right|_{u=0},\quad q\geq p.
\end{equation}
Consequently, (\ref{eq:CPN_intDistributionDelta0}) is
\begin{equation}
\mathcal{I}^{(\delta)}_{\mathfrak{m},a}=\begin{cases}
\frac{i\pi\,\text{sgn}(\mathfrak{m})}{\left(q-p\right)!}\left.\partial^{q-p}\left(e^{-au^{2}}\right)\right|_{u=0}, & q\geq p,\\
0, & q<p.
\end{cases}
\end{equation}
Removing the regulator by taking the $a\to0^{+}$ limit, we find
\begin{equation}
\mathcal{I}^{(\delta)}_{\mathfrak{m}}=\lim_{a\to0^{+}}\mathcal{I}^{(\delta)}_{\mathfrak{m},a}=i\pi\,\text{sgn}(\mathfrak{m})\delta_{q,p}.\label{eq:CPN_intDistributionDelta2}
\end{equation}
Note that this contribution does not require the Gaussian regulator,
since $\left\langle 1,u^{p}\partial^{q}\delta(u)\right\rangle =\delta_{q,p}$
for $p,q\in\mathbb{Z}_{\geq0}$. Next, we evaluate the $1/u$ contribution
to (\ref{eq:CPN_intDistributionReg}), which is 
\begin{equation}
\mathcal{I}^{(1/u)}_{\mathfrak{m},a}=\frac{\left(-1\right)^{q}}{\Gamma\left(q+1\right)}\int_{\mathbb{R}}\text{d}u\left(e^{-au^{2}}u^{p}\partial^{q}\text{pf}\left(\frac{1}{u}\right)\right),\qquad p,q\in\mathbb{Z}_{\geq0}.\label{eq:CPN_intDistributionPV0}
\end{equation}
Applying the distributional identities \cite{kanwal2012generalized}
\begin{equation}
u^{p}\,\partial^{q}\,\text{pf}\left(\frac{1}{u}\right)=\left(-1\right)^{q}q!\,u^{p}\,\text{pf}\left(\frac{1}{u^{q+1}}\right)=\begin{cases}
\left(-1\right)^{q}\,q!\,\text{pf}\left(\frac{1}{u^{q-p+1}}\right), & q\geq p,\\
\left(-1\right)^{q}\,q!\,u^{p-q-1}, & q<p,
\end{cases}\label{eq:CPN_intDistributionPV1}
\end{equation}
to (\ref{eq:CPN_intDistributionPV0}), we obtain 
\begin{equation}
\mathcal{I}^{(1/u)}_{\mathfrak{m},a}=\begin{cases}
\int_{\mathbb{R}}\text{d}u\left(e^{-au^{2}}\text{pf}\left(\frac{1}{u^{q-p+1}}\right)\right), & q\geq p,\\
\int_{\mathbb{R}}\text{d}u\left(e^{-au^{2}}u^{p-q-1}\right), & q<p.
\end{cases}\label{eq:CPN_intDistributionPV2}
\end{equation}
The first of these integrals is evaluated in appendix \ref{appendixA},
where we find 
\begin{equation}
\int_{\mathbb{R}}\text{d}u\left(e^{-au^{2}}\text{pf}\left(\frac{1}{u^{q-p+1}}\right)\right)=\begin{cases}
0, & \text{even }q-p>0,\\
\Gamma\left(\frac{p-q}{2}\right)a^{\frac{q-p}{2}}, & \text{odd }\;q-p>0.
\end{cases}\label{eq:CPN_intDistributionPV3}
\end{equation}
Removing the regulator by taking the $a\to0^{+}$ limit, we have 
\begin{equation}
\mathcal{I}^{(1/u)}_{\mathfrak{m}}=\lim_{a\to0^{+}}\mathcal{I}^{(1/u)}_{\mathfrak{m},a}=0,\quad q\geq p.
\end{equation}
The second integral in (\ref{eq:CPN_intDistributionPV2}) evaluates
to
\begin{equation}
\int_{\mathbb{R}}\text{d}u\left(e^{-au^{2}}u^{p-q-1}\right)=\begin{cases}
0, & \text{even }q-p<0,\\
\Gamma\left(\frac{p-q}{2}\right)a^{\frac{q-p}{2}}, & \text{odd }\;q-p<0.
\end{cases}\label{eq:CPN_intDistributionPV4}
\end{equation}
In the final case, the removal of the Gaussian regulator is obstructed
by a divergence of the form $a^{-k/2},k\in\mathbb{Z}_{>0}$ as $a\to0^{+}$.
To treat this, we remove the Gaussian regulator immediately before
integrating in the $p>q$ regime, then regularize the divergent integral
in a different manner. In particular, we set $a=0$ before evaluating
the integral in (\ref{eq:CPN_intDistributionPV2}), obtaining 
\begin{equation}
\mathcal{I}^{(1/u)}_{\mathfrak{m},a=0}=\int_{\mathbb{R}}u^{p-q-1}\text{d}u=2\int^{\infty}_{0}u^{p-q-1}\text{d}u,\quad p-q>0.\label{eq:CPN_intDistributionPV5}
\end{equation}
The final integral in (\ref{eq:CPN_intDistributionPV5}) is regularized
using an analog of the Hadamard finite-part prescription, denoted
$\text{FP}^{\prime}$, which systematically assigns finite values
to integrals with divergences at infinity \cite{Jones1996}. As described
in appendix \ref{appendixB}, this regularization yields
\begin{equation}
\text{FP}^{\prime}\int^{\infty}_{0}u^{p-q-1}\text{d}u=0,\quad p-q>0,\label{eq:CPN_intDistributionPV6}
\end{equation}
in the sense of equation 51 in \cite{Jones1996}. 

Collecting the contributions (\ref{eq:CPN_intDistributionDelta0})
and (\ref{eq:CPN_intDistributionPV0}) to the integral (\ref{eq:CPN_intDistribution}),
we have
\begin{equation}
\mathcal{I}_{\mathfrak{m}}=\mathcal{I}^{(\delta)}_{\mathfrak{m}}+\mathcal{I}^{(1/u)}_{\mathfrak{m}}=i\pi\,\text{sgn}(\mathfrak{m})\,\delta_{p,q},\quad p,q\in\mathbb{Z}_{\geq0}.\label{eq:CPN_intDistribution_result}
\end{equation}
Accordingly, the correlator (\ref{eq:CPN_correlatorDist}) is
\begin{equation}
\left\langle \prod^{s}_{a=1}\Sigma_{a}(x_{a})\right\rangle =\sum_{\mathfrak{m}\in\mathbb{Z}\backslash\mathsf{M}}\left\langle \prod^{s}_{a=1}\Sigma_{a}(x_{a})\right\rangle _{(\mathfrak{m})}+\sum_{\mathfrak{m}\in\mathsf{M}}\xi^{\mathfrak{m}}\mathcal{I}_{\mathfrak{m}}\label{eq:CPN_correlatorDist2}
\end{equation}
where $\mathsf{M}$ is the set of fluxes in (\ref{eq:CPN_subset}).
By (\ref{eq:CPN_intDistribution_result}), the first sum evaluates
to 
\begin{equation}
\sum_{\mathfrak{m}\in\mathsf{M}}\xi^{\mathfrak{m}}\mathcal{I}_{\mathfrak{m}}=i\pi\sum_{\mathfrak{m}\in\mathsf{M}}\xi^{\mathfrak{m}}\,\text{sgn}(\mathfrak{m})\,\delta_{1,N(\mathfrak{m}+1)-s}=\begin{cases}
i\pi\xi^{\frac{s-N+1}{N}}, & \mathfrak{m}=\frac{s-N+1}{N},\\
0, & \text{otherwise},
\end{cases}\label{eq:CPN_intDistribution_result1}
\end{equation}
where the Kronecker delta imposes the selection rule. This agrees
with the established correlator (\ref{eq:CPN_correlatorJK1}) for
fluxes $\mathfrak{m}\in\mathsf{M}$, up to a factor of $i\pi$. For
example, both (\ref{eq:CPN_correlatorJK1}) and (\ref{eq:CPN_correlatorDist2})
describe the non-zero correlator at $(\mathfrak{m},N,s)=(1,2,3)$. 

To summarize, the selection rule originates from the $\delta(u)$
contribution (\ref{eq:CPN_intDistributionDelta0}), whereas the $1/u$
contribution (\ref{eq:CPN_intDistributionPV0}) vanishes. No regularization
was necessary for the $\delta(u)$ contribution, while the $1/u$
contribution required a Gaussian regulator together with finite-part
prescriptions for divergences at zero and infinity. This regularization
was chosen out of convenience, and the same results are likely attainable
in a number of ways. For instance, the examples of analytic continuation
regularization on page 89 of \cite{kanwal2012generalized} and page
70 of \cite{gel2016generalized} appear to reproduce the results in
equations (\ref{eq:CPN_intDistributionPV3}) and (\ref{eq:CPN_intDistributionPV6}),
respectively. We also expect the same results to follow when combining
the $\delta(u)$ and $1/u$ contributions via the Sokhotski--Plemelj
formula. 

\subsection{Equivalence of descriptions via hyperfunctions\label{subsec:EquivalenceDescriptions}}

In this section, we evaluate the correlator of the $A$-twisted $\mathbb{CP}^{N-1}$
model (\ref{eq:CPN_correlatorDist}) in the framework of hyperfunction
theory \cite{Graf2010,Imai1992} and recover a complex contour integral
description of the correlator that coincides with the established
result (\ref{eq:CPN_correlatorJK1}). This proceeds by first reviewing
a few basic aspects of hyperfunction theory, then repeating the evaluation
of the integral (\ref{eq:CPN_intDistribution}) in terms of hyperfunctions. 

To begin, recall that hyperfunctions and distributions describe a
generalized function $f(x)$ in two different manners. In distribution
theory, $f(x)$ is regarded as a continuous linear functional over
a space of test functions. In hyperfunction theory, $f(x)$ is interpreted
as a difference $F(x+i0)-F(x-i0)$, where $F(x\pm i0)$ are the boundary
values of a function $F(z)$ that is holomorphic in the upper and
lower halves of the complex plane. This notion is already implicit
in the Sokhotski--Plemelj formulas, which equate
\begin{equation}
\delta(x)=-\frac{1}{2\pi i}\left(\frac{1}{x+i0}-\frac{1}{x-i0}\right),\quad\text{pv}\left(\frac{1}{x}\right)=\frac{1}{2}\left(\frac{1}{x+i0}+\frac{1}{x-i0}\right).\label{eq:Hyper1}
\end{equation}
More concretely, hyperfunctions are defined as follows \cite{Graf2010,Imai1992}.
Let $I$ be an interval on the real line $\mathbb{R}$ in $\mathbb{C}$,
let $\mathbb{C}_{\pm}=\left\{ \,z\in\mathbb{C}\,\vert\,\pm\text{Im}\,z>0\right\} $,
let $\mathcal{D}(I)$ be a complex neighborhood of $I$, let $\mathcal{D}_{\pm}(I)=\mathcal{D}(I)\cap\mathbb{C}_{\pm}$,
and let $\mathcal{O}\left(\mathcal{D}(I)\right)$ and $\mathcal{O}\left(\mathcal{D}(I)\backslash I\right)$
be the rings of all holomorphic functions in $\mathcal{D}(I)$ and
$\mathcal{D}(I)\backslash I$, respectively. A hyperfunction on $I$
is defined as an element of the quotient space $\mathcal{B}(I)=\mathcal{O}\left(\mathcal{D}(I)\backslash I\right)/\mathcal{O}\left(\mathcal{D}(I)\right)$,
that is, the quotient space of all holomorphic functions in $\mathcal{D}(I)\backslash I$
over the space of all holomorphic functions in $\mathcal{D}(I)$.
The elements of $\mathcal{B}(I)$ are equivalence classes $\left[F(z)\right]$
with $F(z)\in\mathcal{O}\left(\mathcal{D}(I)\backslash I\right)$.
A hyperfunction on $I$ may be expressed as
\begin{equation}
f(x)=\left[F(z)\right]=F(x+i0)-F(x-i0),\quad F(z)\in\mathcal{O}\left(\mathcal{D}(I)\backslash I\right),\label{eq:Hyper2}
\end{equation}
or equivalently, 
\begin{equation}
f(x)=\left[F_{+}(z),\,F_{-}(z)\right]=F_{+}(x+i0)-F_{-}(x-i0),\quad F_{\pm}(z)\in\mathcal{O}\left(\mathcal{D}_{\pm}(I)\right).\label{eq:Hyper3}
\end{equation}
The Dirac delta and finite part hyperfunctions are defined by 
\begin{equation}
\delta(x)=-\frac{1}{2\pi i}\left[\frac{1}{z},\frac{1}{z}\right],\qquad\text{fp}\left(\frac{1}{x}\right)=\frac{1}{2}\left[\frac{1}{z},-\frac{1}{z}\right],\label{eq:Hyper4}
\end{equation}
respectively \cite{Graf2010}. Hyperfunctions may be differentiated
and multiplied by a real analytic function $\phi(x)$ according to
\begin{equation}
\frac{\text{d}^{n}}{\text{d}x^{n}}f(x)=\left[\frac{\text{d}^{n}}{\text{d}z^{n}}F_{+}(z),\,\frac{\text{d}^{n}}{\text{d}z^{n}}F_{-}(z)\right],\qquad\phi(x)\,f(x)=\left[\phi(z)\,F_{+}(z),\,\phi(z)\,F_{-}(z)\right].\label{eq:Hyper5}
\end{equation}
For example,
\begin{eqnarray}
x^{m}\frac{\text{d}^{n}}{\text{d}x^{n}}\delta(x) & = & -\frac{\left(-1\right)^{n}n!}{2\pi i}\left[\frac{1}{z^{n-m+1}},\frac{1}{z^{n-m+1}}\right],\label{eq:Hyper6}\\
x^{m}\frac{\text{d}^{n}}{\text{d}x^{n}}\text{fp}\left(\frac{1}{x}\right) & = & \frac{\left(-1\right)^{n}n!}{2}\left[\frac{1}{z^{n-m+1}},-\frac{1}{z^{n-m+1}}\right],\label{eq:Hyper7}
\end{eqnarray}
for $m,n\in\mathbb{Z}_{\geq0}$. Given a hyperfunction $f(x)=[F(z)]=[F_{+}(z),\,F_{-}(z)]$
that is holomorphic at both endpoints of a finite interval $[a,b]$
in $\mathbb{R}$, the integral of $f(x)$ over $[a,b]$ is defined
by
\begin{equation}
\int^{b}_{a}f(x)\text{d}x=\int_{\gamma^{+}_{a,b}}F_{+}(z)\text{d}z-\int_{\gamma^{-}_{a,b}}F_{-}(z)\text{d}z=-\oint_{\gamma_{(a,b)}}F(z)\text{d}z,\label{eq:Hyper8}
\end{equation}
where $\gamma^{+}_{a,b}$ goes from $a$ to $b$ through $\mathcal{D}_{+}(I)$,
$\gamma^{-}_{a,b}$ goes from $a$ to $b$ through $\mathcal{D}_{-}(I)$,
and $\gamma_{(a,b)}=\gamma^{-}_{a,b}-\gamma^{+}_{a,b}$ \cite{Graf2010}.
To illustrate (\ref{eq:Hyper8}), let $a<0<b$, then the integral
of $\delta(x)$ is 
\begin{equation}
\int^{b}_{a}\delta(x)\text{d}x=-\frac{1}{2\pi i}\left(\int_{\gamma^{+}_{a,b}}-\int_{\gamma^{-}_{a,b}}\right)\frac{\text{d}z}{z}=\frac{1}{2\pi i}\oint_{\gamma_{(a,b)}}\frac{\text{d}z}{z}=\text{Res}_{z=0}\left(\frac{1}{z}\right)=1,\label{eq:Hyper9}
\end{equation}
while the integral of $\text{fp}(1/x^{2})$ is
\begin{equation}
\int^{b}_{a}\text{fp}\left(\frac{1}{x^{2}}\right)\text{d}x=\frac{1}{2}\left(\int_{\gamma^{+}_{a,b}}+\int_{\gamma^{-}_{a,b}}\right)\frac{\text{d}z}{z^{2}}=-\frac{1}{b}+\frac{1}{a}.\label{eq:Hyper10}
\end{equation}
Observe that the final expression in (\ref{eq:Hyper10}) is the Hadamard finite
part \cite{Imai1992}, in agreement with (\ref{eq:appA_2b}). Since
both $\delta^{(n)}(x)$ and $\text{fp}(1/x^{n+1})$ are holomorphic
in $\mathbb{R}\backslash\left\{ 0\right\} $ for $n\in\mathbb{Z}_{\geq0}$
\cite{Graf2010}, the $a\to-\infty$, $b\to\infty$ limits of (\ref{eq:Hyper9})
and (\ref{eq:Hyper10}) exist. 

Returning to the \emph{A}-twisted $\mathbb{CP}^{N-1}$ correlator,
we evaluate the integral (\ref{eq:CPN_intDistribution}) in the framework
of hyperfunctions. In particular, we consider 
\begin{equation}
\mathcal{I}_{\mathfrak{m}}=\frac{\left(-1\right)^{q}}{q!}\int^{\infty}_{-\infty}\text{d}u\left(u^{p}\partial^{q}\left(\text{fp}\left(\frac{1}{u}\right)+i\pi\,\text{sgn}(\mathfrak{m})\,\delta(u)\right)\right),\quad p,q\in\mathbb{Z}_{\geq0},\label{eq:Hyper11}
\end{equation}
where $\text{fp}\left(1/u\right)$ and $\delta(u)$ are the hyperfunctions
defined in (\ref{eq:Hyper4}). Differentiation and multiplication
by $u^{p}$ follows the examples in (\ref{eq:Hyper6}) and (\ref{eq:Hyper7}),
and the resulting hyperfunctions are holomorphic in $\mathbb{R}\backslash\left\{ 0\right\} $.
By the first equality in (\ref{eq:Hyper8}), the integral (\ref{eq:Hyper11})
is
\begin{equation}
\mathcal{I}_{\mathfrak{m}}=\frac{1}{2}\left(\int_{\gamma^{+}_{-\infty,\infty}}+\int_{\gamma^{-}_{-\infty,\infty}}\right)\frac{\text{d}z}{z^{q-p+1}}-\frac{i\pi\,\text{sgn}(\mathfrak{m})}{2\pi i}\left(\int_{\gamma^{\prime+}_{-\infty,\infty}}-\int_{\gamma^{\prime-}_{-\infty,\infty}}\right)\frac{\text{d}z}{z^{q-p+1}}.\label{eq:Hyper12}
\end{equation}
The first two integrals contribute a vanishing finite part\footnote{For $q\geq p$, the finite part vanishes by the arguments of appendix \ref{appendixA}. For $p>q$, the finite part vanishes by the arguments of appendix \ref{appendixB}.}, while the last two integrals combine according to the final equality
in (\ref{eq:Hyper8}). Accordingly, (\ref{eq:Hyper12}) reduces to
\begin{equation}
\mathcal{I}_{\mathfrak{m}}=\frac{i\pi\,\text{sgn}(\mathfrak{m})}{2\pi i}\oint_{\gamma^{\prime}_{(-\infty,\infty)}}\frac{\text{d}z}{z^{q-p+1}},\label{eq:Hyper13}
\end{equation}
where the contour $\gamma^{\prime}_{(-\infty,\infty)}=\gamma^{\prime-}_{-\infty,\infty}-\gamma^{\prime+}_{-\infty,\infty}$
is closed and positively oriented. Evaluating (\ref{eq:Hyper13}),
we find 
\begin{equation}
\mathcal{I}_{\mathfrak{m}}=i\pi\,\text{sgn}(\mathfrak{m})\text{Res}_{z=0}\left(\frac{1}{z^{q-p+1}}\right)=i\pi\,\text{sgn}(\mathfrak{m})\delta_{q,p},\label{eq:Hyper14}
\end{equation}
in agreement with distributional result (\ref{eq:CPN_intDistribution_result}).
In view of (\ref{eq:Hyper13}), the correlator (\ref{eq:CPN_correlatorDist})
is
\begin{equation}
\left\langle \prod^{s}_{a=1}\Sigma_{a}(x_{a})\right\rangle =\sum_{\mathfrak{m}\in\mathbb{Z}\backslash\mathsf{M}}\left\langle \prod^{s}_{a=1}\Sigma_{a}(x_{a})\right\rangle _{(\mathfrak{m})}+i\pi\sum_{\mathfrak{m}\in\mathsf{M}}\oint_{\gamma^{\prime}_{(-\infty,\infty)}}\frac{\text{d}z}{2\pi i}\frac{\xi^{\mathfrak{m}}}{z^{N(\mathfrak{m}+1)-s}},\label{eq:Hyper15}
\end{equation}
where $\mathsf{M}$ is the set of fluxes in (\ref{eq:CPN_subset}). 

The correlator (\ref{eq:Hyper15}) agrees with the established complex
contour integral description in (\ref{eq:CPN_correlatorJK1}) for
fluxes $\mathfrak{m}\in\mathsf{M}$, up to a factor of $i\pi$. In
particular, the contour $\gamma^{\prime}_{(-\infty,\infty)}$ coincides
with the one specified by the JK residue prescription when the JK
parameter is $\eta>0$. In contrast with the approach of section (\ref{subsec:DistributionalDescription}),
the hyperfunction interpretation of the correlator (\ref{eq:CPN_correlatorDist})
required no test function and yielded the appropriate residue directly.
For these reasons, the hyperfunction framework appears better suited
to the present computation. 

\section{Discussion \label{sec:Discussion}}

In this paper, we applied a stationary phase version of supersymmetric
localization to topologically $A$-twisted $\mathcal{N}=(2,2)$ supersymmetric
gauged linear sigma models (GLSMs) on $S^{2}$. Our primary result
is a novel and exact formula for observables of abelian $A$-twisted
$\mathcal{N}=(2,2)$ GLSMs on $S^{2}$, described by a distribution
integrated along the real line. We verified the distributional integral
formula by computing the correlator of the $A$-twisted $\mathbb{CP}^{N-1}$
model, reproducing the known selection rule. Finally, hyperfunctions
were used to relate the distributional integral description of the
$\mathbb{CP}^{N-1}$ correlator to the established complex contour
integral description obtained via localization in \cite{Benini:2015noa,Closset:2015rna,Benini:2016hjo}. 

This work is a tangible step to reconciling discrepancies between
two different approaches to supersymmetric localization. On the one
hand, the Jeffrey-Kirwan (JK) residue prescription was introduced
in \cite{JeffreyKirwan} and extended to the context of supersymmetric
localization in \cite{Benini:2013nda}. Supersymmetric localization
techniques employing the Jeffrey-Kirwan (JK) residue prescription
were developed in \cite{Benini:2013xpa,Cordova:2014oxa,Ohta:2019odi,Closset:2015ohf,Closset:2017bse,Closset:2017zgf,Closset:2019hyt,Benini:2015noa,Hosseini:2018uzp,Benini:2016hjo,Hori:2014tda},
and applied to $A$-twisted $\mathcal{N}=(2,2)$ GLSMs on $S^{2}$
in \cite{Benini:2015noa,Closset:2015rna,Benini:2016hjo}. This approach
to localization leads to observables described by JK residues, or
equivalently, a meromorphic function integrated along a complex contour
specified by the JK residue prescription. On the other hand, the work
presented here builds on the non-abelian localization techniques that
were introduced in \cite{Witten:1992xu} and developed in \cite{Leeb-Lundberg:2023jsj,Griguolo:2024ecw}.
Specifically, we utilize the stationary phase version of supersymmetric
localization introduced in \cite{Griguolo:2024ecw}, which leads to
observables described by a distribution integrated along a real contour.
Our localization computation differs from those of \cite{Benini:2015noa,Ohta:2019odi,Closset:2015rna,Benini:2016hjo}
by a $Q$-exact term, in the sense that we take $t>0$ and $\tau=0$
in the localizing term (\ref{eq:locTerm2_QVvec}) instead of
$t=0$ and $\tau=1$. Consequently, the two approaches lead to mathematically
distinct descriptions of observables that were expected to be equivalent.
The work presented here provides the first concrete evidence of this
equivalence in the well-studied setting of $A$-twisted $\mathcal{N}=(2,2)$
GLSMs. 

The bridge between the distributional and JK residue descriptions
of observables is provided by hyperfunction theory \cite{Graf2010,Imai1992}.
We recovered the complex contour integral description of the $A$-twisted
$\mathbb{CP}^{N-1}$ correlator from the distributional integral description
by interpreting the distributions as hyperfunctions. Specifically,
our contour encircles a pole at the origin in the complex plane of
the continuous modulus $u$. The JK residue, on the other hand, captures
the contribution of an analogous pole at the origin in the $u$-plane
when the JK parameter is taken to be positive. In this sense, hyperfunction
theory offers a framework for relating distributional and JK residue
descriptions of observables arising from localization computations
that differ by $Q$-exact terms.

Our derivation hinges on the evaluation of the distributional one-loop
contribution. The integrals over fluctuations were evaluated using
monopole spherical harmonics on $S^{2}$. Crucially, we identified
a bosonic scalar mode with purely imaginary eigenvalue, associated
to an oscillatory integral lacking a  positive definite quadratic
form. This oscillatory integral was evaluated as a distribution, while
the remaining integrals over fluctuations were evaluated as determinants.
Consequently, the final one-loop contribution is a distribution multiplied
by a ratio of fluctuation determinants. This work
builds on the techniques introduced in \cite{Griguolo:2024ecw}, where
a distributional one-loop contribution was evaluated using an index
theorem. 

The stationary phase localization techniques presented here likely
generalize to other settings. As pointed out in \cite{Griguolo:2024ecw},
such extensions may prove useful in addressing unsatisfactory aspects
of supersymmetric localization techniques employing JK residue type
prescriptions. In particular, the JK prescription appears to apply
only to gauge theories with charged matter. It requires a priori knowledge
of the charges of the dynamical matter multiplets, and the charge
data must satisfy specific conditions. These requirements likely obstruct
the weak gauging of global symmetries at the level of the localized
partition function, as noted in \cite{Griguolo:2024ecw}. Although
extensions to theories lacking the requisite charge data are conceivable
\cite{Benini:2016hjo,Closset:2015rna}, they require ad hoc modifications
of the JK prescription, whereas alternative localization techniques
offer conceptual and technical advantages \cite{Leeb-Lundberg:2023jsj}.
Several localization computations that use JK residue type prescriptions
are promising candidates for generalizing stationary phase localization
techniques. Examples include $\Omega$-deformed $\mathcal{N}=(2,2)$
GLSMs on $S^{2}$ \cite{Closset:2015rna}, $A/2$-twisted $\mathcal{N}=(0,2)$
GLSMs on $S^{2}$ \cite{Closset:2015ohf}, four-dimensional twisted
$\mathcal{N}=2$ theories on toric Kähler manifolds \cite{Bershtein:2015xfa},
and five-dimensional $\mathcal{N}=1$ indices \cite{Hosseini:2018uzp}.
A key aspect of any such generalization would be to demonstrate, via
hyperfunction theory, the equivalence between the distributional and
complex contour integral descriptions of observables.

In upcoming work, we consider extensions to non-abelian gauge groups
and $\Omega$-deformed $\mathcal{N}=(2,2)$ GLSMs. The former extension
would begin with gauge group $\text{SU}(2)$, and likely leverage
the insights from the localization computations involving non-abelian
gauge groups in \cite{Leeb-Lundberg:2023jsj,Griguolo:2024ecw}. After
deriving a distributional integral description of observables for
non-abelian $A$-twisted $\mathcal{N}=(2,2)$ GLSMs, one could attempt
to recover the partition function as a sum over solutions of the Bethe
ansatz equations \cite{Benini:2016hjo,Nekrasov:2014xaa}. Applying
a stationary phase version of supersymmetric localization to $\Omega$-deformed
$\mathcal{N}=(2,2)$ GLSMs on $S^{2}$ would complement the results
of \cite{Closset:2015rna}, and the key point would be to ensure the
$\Omega$-deformation parameter does not spoil any of the arguments
leading to a distributional integral description of observables. In
both cases, we aim to verify the localization computation by evaluating
at least one distributional observable, then relating the distributional
and contour integral descriptions of observables through hyperfunctions.

\acknowledgments
I thank Itamar Yaakov for valuable insights and suggestions. I also thank Can Kozçaz and Paolo Di Vecchia for helpful discussions. I am grateful to Charlotte Kristjansen and the other faculty members at the Niels Bohr Institute at Copenhagen University for their hospitality. The work was supported by the Scientific and Technological Research Council of Turkey (TÜBİTAK) under the Grant Number 124F428. ELL thanks TÜBİTAK for their support. This article is based upon work from COST Action 22113 THEORY-CHALLENGES, supported by COST (European Cooperation in Science and Technology).
\newpage
\appendix

\section{Hadamard finite part} \label{appendixA}

In this appendix, we evaluate the integral in (\ref{eq:CPN_intDistributionPV3}).
As described in chapter 4 of \cite{kanwal2012generalized}, the distribution
$\text{pf}\left(\frac{1}{x^{n+1}}\right)$ is defined by 
\begin{eqnarray}
\left\langle \text{pf}\left(\frac{1}{x^{n+1}}\right),\varphi(x)\right\rangle  & = & \text{FP}\int_{\mathbb{R}}\frac{\varphi(x)-\left(\sum^{n}_{k=0}\frac{\varphi^{(k)}(0)}{k!}x^{k}\right)\Theta\left(1-\frac{\left|x\right|}{\varepsilon}\right)}{x^{n+1}}\text{d}x\label{eq:appA_1}\\
 & = & \text{FP}\left(\int_{\left|x\right|<\varepsilon}\frac{\varphi(x)-\sum^{n}_{k=0}\frac{x^{k}}{k!}\varphi^{(k)}(0)}{x^{n+1}}\text{d}x+\int_{\left|x\right|>\varepsilon}\frac{\varphi(x)}{x^{n+1}}\text{d}x\right)
\end{eqnarray}
for $n\in\mathbb{Z}_{\geq0}$. Here $\varphi(x)$ is a test function
with compact support or rapid decay, $\Theta(x)$ is the Heaviside
step function, and $\text{FP}$ denotes the Hadamard finite part prescription.
The finite part prescription systematically assigns values to integrals
with divergences at finite points. For example, the integral
\begin{equation}
\int^{b}_{a}\frac{\text{d}x}{x^{2}},\qquad a<0<b,\label{eq:appA_2}
\end{equation}
is divergent at $x=0$ in the conventional sense, but its finite part
is 
\begin{eqnarray}
\text{FP}\int^{b}_{a}\frac{\text{d}x}{x^{2}} & = & \text{FP}\left(\int^{-\varepsilon}_{a}\frac{\text{d}x}{x^{2}}+\int^{b}_{\varepsilon}\frac{\text{d}x}{x^{2}}\right)\\
 & = & \text{FP}\left(-\frac{1}{b}+\frac{1}{a}+\frac{2}{\varepsilon}\right)\\
 & = & -\frac{1}{b}+\frac{1}{a}.\label{eq:appA_2b}
\end{eqnarray}
The $\text{FP}$ operation discards inverse powers of $\varepsilon$
and logarithms that diverge as $\varepsilon\to0$. Notice that the
Cauchy principal value of (\ref{eq:appA_2}) does not exist due to
the divergence of $2/\varepsilon$ as $\varepsilon\to0$. 

The integral in (\ref{eq:CPN_intDistributionPV3}) is described by
(\ref{eq:appA_1}) with $\varphi(x)=e^{-ax^{2}},\,a>0$ and $n=q-p$.
We evaluate the first few values of $n$ explicitly, then describe
the general result. At $n=0$, we have 
\begin{eqnarray}
\left\langle \text{pf}\left(\frac{1}{x}\right),e^{-ax^{2}}\right\rangle  & = & \text{FP}\left(\int_{\left|x\right|<\varepsilon}\frac{e^{-ax^{2}}-1}{x}\text{d}x+\int_{\left|x\right|>\varepsilon}\frac{e^{-ax^{2}}}{x}\text{d}x\right).\label{eq:appA_3}
\end{eqnarray}
The integrals evaluate to 
\begin{eqnarray}
\int^{\varepsilon}_{-\varepsilon}\frac{e^{-ax^{2}}-1}{x}\text{d}x & = & 0,\\
\int^{\infty}_{\varepsilon}\frac{e^{-ax^{2}}}{x}\text{d}x & = & -\int^{-\varepsilon}_{-\infty}\frac{e^{-ax^{2}}}{x}\text{d}x=\frac{\Gamma\left(0,a\varepsilon^{2}\right)}{2},
\end{eqnarray}
where $\Gamma(b,y)=\int^{\infty}_{y}t^{b-1}e^{-t}\text{d}t$ is the
upper incomplete gamma function. Accordingly, (\ref{eq:appA_3}) is
\begin{equation}
\left\langle \text{pf}\left(\frac{1}{x}\right),e^{-ax^{2}}\right\rangle =\text{FP}\left(0\right)=0.
\end{equation}
and the finite part coincides with the principal value. The $n\in2\mathbb{Z}_{>0}$
cases are essentially more complicated versions of the $n=0$ case,
in the sense that the finite part coincides with the principal value,
and the integral evaluates to zero. Specifically, one finds 
\begin{equation}
\left\langle \text{pf}\left(\frac{1}{x^{2k+1}}\right),e^{-ax^{2}}\right\rangle =0,\qquad k=0,1,2,\dots.\label{eq:appA_4}
\end{equation}
In contrast, when $n$ is an odd positive integer, the principal value
does not exist and the finite part is non-vanishing. At $n=1$, we
have 
\begin{equation}
\left\langle \text{pf}\left(\frac{1}{x^{2}}\right),e^{-ax^{2}}\right\rangle =\text{FP}\left(\int_{\left|x\right|<\varepsilon}\frac{e^{-ax^{2}}-1}{x^{2}}\text{d}x+\int_{\left|x\right|>\varepsilon}\frac{e^{-ax^{2}}}{x^{2}}\text{d}x\right).\label{eq:appA_5}
\end{equation}
The integrals evaluate to 
\begin{eqnarray}
\int^{\varepsilon}_{-\varepsilon}\frac{e^{-ax^{2}}-1}{x^{2}}\text{d}x & = & \frac{2}{\varepsilon}-\frac{2e^{-a\varepsilon^{2}}}{\varepsilon}-2\sqrt{\pi}\sqrt{a}\text{erf}\left(\sqrt{a}\varepsilon\right),\\
\int_{\left|x\right|>\varepsilon}\frac{e^{-ax^{2}}}{x^{2}}\text{d}x & = & \frac{2e^{-a\varepsilon^{2}}}{\varepsilon}-2\sqrt{\pi}\sqrt{a}\text{erfc}\left(\sqrt{a}\varepsilon\right)
\end{eqnarray}
where $\text{erf}(y)=\frac{2}{\sqrt{\pi}}\int^{y}_{0}e^{-t^{2}}\text{d}t$
is the error function and $\text{erfc}(y)=1-\text{erf}(y)$. The error
functions cancel, and (\ref{eq:appA_5}) reduces to 
\begin{equation}
\left\langle \text{pf}\left(\frac{1}{x^{2}}\right),e^{-ax^{2}}\right\rangle =\text{FP}\left(\frac{2}{\varepsilon}-2\sqrt{\pi}\sqrt{a}\right)=2\sqrt{\pi}a.
\end{equation}
At $n=3$, we have 
\begin{eqnarray}
\left\langle \text{pf}\left(\frac{1}{x^{4}}\right),e^{-ax^{2}}\right\rangle  & = & \text{FP}\left(\int_{\left|x\right|<\varepsilon}\frac{e^{-ax^{2}}-1+ax^{2}}{x^{4}}\text{d}x+\int_{\left|x\right|>\varepsilon}\frac{e^{-ax^{2}}}{x^{4}}\text{d}x\right).\label{eq:appA_6}
\end{eqnarray}
The integrals evaluate to 
\begin{eqnarray}
\int^{\varepsilon}_{-\varepsilon}\frac{e^{-ax^{2}}-1+ax^{2}}{x^{4}}\text{d}x & = & \frac{2\left(1-e^{-a\varepsilon^{2}}\right)}{3\varepsilon^{3}}+\frac{2\left(2ae^{-a\varepsilon^{2}}-3a\right)}{3\varepsilon}+\frac{4}{3}\sqrt{\pi}a^{3/2}\text{erf}\left(\sqrt{a}\varepsilon\right),\\
\int_{\left|x\right|>\varepsilon}\frac{e^{-ax^{2}}}{x^{4}}\text{d}x & = & \frac{2e^{-a\varepsilon^{2}}}{3\varepsilon^{3}}-\frac{4ae^{-a\varepsilon^{2}}}{3\varepsilon}+\frac{4}{3}\sqrt{\pi}a^{3/2}\text{erfc}\left(\sqrt{a}\varepsilon\right).
\end{eqnarray}
The error functions cancel, such that (\ref{eq:appA_6}) is
\begin{equation}
\left\langle \text{pf}\left(\frac{1}{x^{4}}\right),e^{-ax^{2}}\right\rangle =\text{FP}\left(\frac{2}{3\varepsilon^{3}}-\frac{2a}{\varepsilon}+\frac{4}{3}\sqrt{\pi}a^{3/2}\right)=\frac{4}{3}\sqrt{\pi}a^{3/2}.
\end{equation}
Continuing in this manner, one finds 
\begin{eqnarray}
\left\langle \text{pf}\left(\frac{1}{x^{2k}}\right),e^{-ax^{2}}\right\rangle  & = & \text{FP}\left(\frac{\Gamma\left(\frac{1}{2}-k\right)}{a^{\frac{1}{2}-k}}+\sum^{k-1}_{m=0}\frac{2\varepsilon^{1-2k}\left(-a\varepsilon^{2}\right)^{m}}{m!(2k-2m-1)}\right)=\frac{\Gamma\left(\frac{1}{2}-k\right)}{a^{\frac{1}{2}-k}}\label{eq:appA_7}
\end{eqnarray}
for $k=1,2,\dots$. In view of (\ref{eq:appA_4}) and (\ref{eq:appA_7}),
we have 
\[
\left\langle \text{pf}\left(\frac{1}{x^{n+1}}\right),e^{-ax^{2}}\right\rangle =\begin{cases}
0, & n=2k,\\
a^{n/2}\Gamma\left(-\frac{n}{2}\right), & n=2k+1,
\end{cases}
\]
for $k\in\mathbb{Z}_{\geq0}$.
Reinstating $n=q-p$, one obtains the expression in (\ref{eq:CPN_intDistributionPV3}).

\section{Finite part at infinity} \label{appendixB}

In this appendix, we evaluate  the integral in (\ref{eq:CPN_intDistributionPV6}).
As described in \cite{Jones1996}, integrals of the form 
\begin{equation}
\int^{\infty}_{0}x^{\beta}\text{d}x,\qquad\beta\in\mathbb{Z}_{\geq0}
\end{equation}
are divergent in the conventional sense, but may be regularized using
an analog of the Hadamard finite-part prescription, denoted $\text{FP}^{\prime}$.
The regularization proceeds by splitting the integral at an arbitrary
finite point $c>0$, yielding
\begin{equation}
\text{FP}^{\prime}\int^{\infty}_{0}x^{\beta}\text{d}x=\int^{c}_{0}x^{\beta}\text{d}x+\int^{\infty}_{c}x^{\beta}\text{d}x.\label{eq:fpInf2}
\end{equation}
The first integral on the right hand side of the equality converges
in the conventional sense 
\begin{equation}
\int^{c}_{0}x^{\beta}\text{d}x=\frac{c^{\beta+1}}{\beta+1}.
\end{equation}
The final integral in (\ref{eq:fpInf2}) is evaluated according to
the formula 
\begin{equation}
\int^{\infty}_{c}\frac{\text{d}g(x)}{\text{d}x}\text{d}x=-g(c)+\text{fin}\,g(\xi).\label{eq:fpInf3}
\end{equation}
Here, $\text{fin}g(\xi)$ denotes the finite part of $g(\xi)$, which
is determined by $g(\xi)$ as $\xi\to\infty$. The $\text{fin}$ operation
discards terms of the form $\xi^{r}\ln^{n}\xi$, e.g. $\text{fin}\left(\xi^{3/2}\ln^{2}\xi+x+1/\xi^{2}\right)=x.$
Setting $g(x)=x^{\beta+1}/(\beta+1)$ in (\ref{eq:fpInf3}), we have
\begin{eqnarray}
\int^{\infty}_{c}x^{\beta}\text{d}x & = & -\frac{c^{\beta+1}}{\beta+1}+\text{fin}\left(\frac{\xi^{\beta+1}}{\beta+1}\right)=-\frac{c^{\beta+1}}{\beta+1}.
\end{eqnarray}
Accordingly, (\ref{eq:fpInf2}) is
\begin{equation}
\text{FP}^{\prime}\int^{\infty}_{0}x^{\beta}\text{d}x=\frac{c^{\beta+1}}{\beta+1}-\frac{c^{\beta+1}}{\beta+1}=0.
\end{equation}
Reinstating $\beta=p-q-1$, one obtains the result in (\ref{eq:CPN_intDistributionPV6}).



\bibliographystyle{JHEP}
\bibliography{glsmsAmodelv2}

\providecommand{\href}[2]{#2}\begingroup\raggedright\begin{thebibliography}{10}

\bibitem{Pestun_2017}
V.~Pestun, M.~Zabzine, F.~Benini, T.~Dimofte, T.T.~Dumitrescu, K.~Hosomichi et~al., \emph{Localization techniques in quantum field theories}, \href{https://doi.org/10.1088/1751-8121/aa63c1}{\emph{Journal of Physics A: Mathematical and Theoretical} {\bfseries 50} (2017) 440301}.

\bibitem{Benini:2015noa}
F.~Benini and A.~Zaffaroni, \emph{{A topologically twisted index for three-dimensional supersymmetric theories}}, \href{https://doi.org/10.1007/JHEP07(2015)127}{\emph{JHEP} {\bfseries 07} (2015) 127} [\href{https://arxiv.org/abs/1504.03698}{{\ttfamily 1504.03698}}].

\bibitem{Closset:2015rna}
C.~Closset, S.~Cremonesi and D.S.~Park, \emph{{The equivariant A-twist and gauged linear sigma models on the two-sphere}}, \href{https://doi.org/10.1007/JHEP06(2015)076}{\emph{JHEP} {\bfseries 06} (2015) 076} [\href{https://arxiv.org/abs/1504.06308}{{\ttfamily 1504.06308}}].

\bibitem{Witten:1993yc}
E.~Witten, \emph{{Phases of N=2 theories in two-dimensions}}, \href{https://doi.org/10.1016/0550-3213(93)90033-L}{\emph{Nucl. Phys. B} {\bfseries 403} (1993) 159} [\href{https://arxiv.org/abs/hep-th/9301042}{{\ttfamily hep-th/9301042}}].

\bibitem{Morrison:1994fr}
D.R.~Morrison and M.R.~Plesser, \emph{{Summing the instantons: Quantum cohomology and mirror symmetry in toric varieties}}, \href{https://doi.org/10.1016/0550-3213(95)00061-V}{\emph{Nucl. Phys. B} {\bfseries 440} (1995) 279} [\href{https://arxiv.org/abs/hep-th/9412236}{{\ttfamily hep-th/9412236}}].

\bibitem{Gaiotto:2015aoa}
D.~Gaiotto, G.W.~Moore and E.~Witten, \emph{{Algebra of the Infrared: String Field Theoretic Structures in Massive ${\cal N}=(2,2)$ Field Theory In Two Dimensions}},  \href{https://arxiv.org/abs/1506.04087}{{\ttfamily 1506.04087}}.

\bibitem{Hori:2006dk}
K.~Hori and D.~Tong, \emph{{Aspects of Non-Abelian Gauge Dynamics in Two-Dimensional N=(2,2) Theories}}, \href{https://doi.org/10.1088/1126-6708/2007/05/079}{\emph{JHEP} {\bfseries 05} (2007) 079} [\href{https://arxiv.org/abs/hep-th/0609032}{{\ttfamily hep-th/0609032}}].

\bibitem{Chen:2020iyo}
Z.~Chen, J.~Guo and M.~Romo, \emph{{A GLSM View on Homological Projective Duality}}, \href{https://doi.org/10.1007/s00220-022-04401-1}{\emph{Commun. Math. Phys.} {\bfseries 394} (2022) 355} [\href{https://arxiv.org/abs/2012.14109}{{\ttfamily 2012.14109}}].

\bibitem{Jockers:2012dk}
H.~Jockers, V.~Kumar, J.M.~Lapan, D.R.~Morrison and M.~Romo, \emph{{Two-Sphere Partition Functions and Gromov-Witten Invariants}}, \href{https://doi.org/10.1007/s00220-013-1874-z}{\emph{Commun. Math. Phys.} {\bfseries 325} (2014) 1139} [\href{https://arxiv.org/abs/1208.6244}{{\ttfamily 1208.6244}}].

\bibitem{Gomis:2012wy}
J.~Gomis and S.~Lee, \emph{{Exact Kahler Potential from Gauge Theory and Mirror Symmetry}}, \href{https://doi.org/10.1007/JHEP04(2013)019}{\emph{JHEP} {\bfseries 04} (2013) 019} [\href{https://arxiv.org/abs/1210.6022}{{\ttfamily 1210.6022}}].

\bibitem{Benini:2013nda}
F.~Benini, R.~Eager, K.~Hori and Y.~Tachikawa, \emph{{Elliptic genera of two-dimensional N=2 gauge theories with rank-one gauge groups}}, \href{https://doi.org/10.1007/s11005-013-0673-y}{\emph{Lett. Math. Phys.} {\bfseries 104} (2014) 465} [\href{https://arxiv.org/abs/1305.0533}{{\ttfamily 1305.0533}}].

\bibitem{Gadde:2013dda}
A.~Gadde and S.~Gukov, \emph{{2d Index and Surface operators}}, \href{https://doi.org/10.1007/JHEP03(2014)080}{\emph{JHEP} {\bfseries 03} (2014) 080} [\href{https://arxiv.org/abs/1305.0266}{{\ttfamily 1305.0266}}].

\bibitem{Halverson:2013qca}
J.~Halverson, H.~Jockers, J.M.~Lapan and D.R.~Morrison, \emph{{Perturbative Corrections to Kaehler Moduli Spaces}}, \href{https://doi.org/10.1007/s00220-014-2157-z}{\emph{Commun. Math. Phys.} {\bfseries 333} (2015) 1563} [\href{https://arxiv.org/abs/1308.2157}{{\ttfamily 1308.2157}}].

\bibitem{Hori:2000kt}
K.~Hori and C.~Vafa, \emph{{Mirror symmetry}},  \href{https://arxiv.org/abs/hep-th/0002222}{{\ttfamily hep-th/0002222}}.

\bibitem{Benini:2012ui}
F.~Benini and S.~Cremonesi, \emph{{Partition Functions of ${\mathcal{N}=(2,2)}$ Gauge Theories on S$^{2}$ and Vortices}}, \href{https://doi.org/10.1007/s00220-014-2112-z}{\emph{Commun. Math. Phys.} {\bfseries 334} (2015) 1483} [\href{https://arxiv.org/abs/1206.2356}{{\ttfamily 1206.2356}}].

\bibitem{Doroud:2012xw}
N.~Doroud, J.~Gomis, B.~Le~Floch and S.~Lee, \emph{{Exact Results in D=2 Supersymmetric Gauge Theories}}, \href{https://doi.org/10.1007/JHEP05(2013)093}{\emph{JHEP} {\bfseries 05} (2013) 093} [\href{https://arxiv.org/abs/1206.2606}{{\ttfamily 1206.2606}}].

\bibitem{Benini:2013xpa}
F.~Benini, R.~Eager, K.~Hori and Y.~Tachikawa, \emph{{Elliptic Genera of 2d ${\mathcal{N}}$ = 2 Gauge Theories}}, \href{https://doi.org/10.1007/s00220-014-2210-y}{\emph{Commun. Math. Phys.} {\bfseries 333} (2015) 1241} [\href{https://arxiv.org/abs/1308.4896}{{\ttfamily 1308.4896}}].

\bibitem{Hori:2013ika}
K.~Hori and M.~Romo, \emph{{Exact Results In Two-Dimensional (2,2) Supersymmetric Gauge Theories With Boundary}},  \href{https://arxiv.org/abs/1308.2438}{{\ttfamily 1308.2438}}.

\bibitem{Honda:2013uca}
D.~Honda and T.~Okuda, \emph{{Exact results for boundaries and domain walls in 2d supersymmetric theories}}, \href{https://doi.org/10.1007/JHEP09(2015)140}{\emph{JHEP} {\bfseries 09} (2015) 140} [\href{https://arxiv.org/abs/1308.2217}{{\ttfamily 1308.2217}}].

\bibitem{Leeb-Lundberg:2023jsj}
E.H.~Leeb-Lundberg, \emph{{Unstable instantons in A-model localization}}, \href{https://doi.org/10.1007/JHEP09(2024)190}{\emph{JHEP} {\bfseries 09} (2024) 190} [\href{https://arxiv.org/abs/2312.11347}{{\ttfamily 2312.11347}}].

\bibitem{Benini:2016hjo}
F.~Benini and A.~Zaffaroni, \emph{{Supersymmetric partition functions on Riemann surfaces}}, {\emph{Proc. Symp. Pure Math.} {\bfseries 96} (2017) 13} [\href{https://arxiv.org/abs/1605.06120}{{\ttfamily 1605.06120}}].

\bibitem{Ohta:2019odi}
K.~Ohta and N.~Sakai, \emph{{Higgs and Coulomb Branch Descriptions of the Volume of the Vortex Moduli Space}}, \href{https://doi.org/10.1093/ptep/ptz016}{\emph{PTEP} {\bfseries 2019} (2019) 043B01} [\href{https://arxiv.org/abs/1811.03824}{{\ttfamily 1811.03824}}].

\bibitem{Hosomichi:2017dbc}
K.~Hosomichi, S.~Lee and T.~Okuda, \emph{{Supersymmetric vortex defects in two dimensions}}, \href{https://doi.org/10.1007/JHEP01(2018)033}{\emph{JHEP} {\bfseries 01} (2018) 033} [\href{https://arxiv.org/abs/1705.10623}{{\ttfamily 1705.10623}}].

\bibitem{Bonelli:2013mma}
G.~Bonelli, A.~Sciarappa, A.~Tanzini and P.~Vasko, \emph{{Vortex partition functions, wall crossing and equivariant Gromov-Witten invariants}}, \href{https://doi.org/10.1007/s00220-014-2193-8}{\emph{Commun. Math. Phys.} {\bfseries 333} (2015) 717} [\href{https://arxiv.org/abs/1307.5997}{{\ttfamily 1307.5997}}].

\bibitem{Gerhardus:2018zwb}
A.~Gerhardus, H.~Jockers and U.~Ninad, \emph{{The Geometry of Gauged Linear Sigma Model Correlation Functions}}, \href{https://doi.org/10.1016/j.nuclphysb.2018.06.008}{\emph{Nucl. Phys. B} {\bfseries 933} (2018) 65} [\href{https://arxiv.org/abs/1803.10253}{{\ttfamily 1803.10253}}].

\bibitem{Closset:2017vvl}
C.~Closset, N.~Mekareeya and D.S.~Park, \emph{{A-twisted correlators and Hori dualities}}, \href{https://doi.org/10.1007/JHEP08(2017)101}{\emph{JHEP} {\bfseries 08} (2017) 101} [\href{https://arxiv.org/abs/1705.04137}{{\ttfamily 1705.04137}}].

\bibitem{Kim:2016jye}
B.~Kim, J.~Oh, K.~Ueda and Y.~Yoshida, \emph{{Residue mirror symmetry for Grassmannians}},  \href{https://arxiv.org/abs/1607.08317}{{\ttfamily 1607.08317}}.

\bibitem{Ueda:2016wfa}
K.~Ueda and Y.~Yoshida, \emph{{Equivariant A-twisted GLSM and Gromov--Witten invariants of CY 3-folds in Grassmannians}}, \href{https://doi.org/10.1007/JHEP09(2017)128}{\emph{JHEP} {\bfseries 09} (2017) 128} [\href{https://arxiv.org/abs/1602.02487}{{\ttfamily 1602.02487}}].

\bibitem{Nekrasov:2009uh}
N.A.~Nekrasov and S.L.~Shatashvili, \emph{{Supersymmetric vacua and Bethe ansatz}}, \href{https://doi.org/10.1016/j.nuclphysbps.2009.07.047}{\emph{Nucl. Phys. B Proc. Suppl.} {\bfseries 192-193} (2009) 91} [\href{https://arxiv.org/abs/0901.4744}{{\ttfamily 0901.4744}}].

\bibitem{Nekrasov:2014xaa}
N.A.~Nekrasov and S.L.~Shatashvili, \emph{{Bethe/Gauge correspondence on curved spaces}}, \href{https://doi.org/10.1007/JHEP01(2015)100}{\emph{JHEP} {\bfseries 01} (2015) 100} [\href{https://arxiv.org/abs/1405.6046}{{\ttfamily 1405.6046}}].

\bibitem{Pestun:2007rz}
V.~Pestun, \emph{{Localization of gauge theory on a four-sphere and supersymmetric Wilson loops}}, \href{https://doi.org/10.1007/s00220-012-1485-0}{\emph{Commun. Math. Phys.} {\bfseries 313} (2012) 71} [\href{https://arxiv.org/abs/0712.2824}{{\ttfamily 0712.2824}}].

\bibitem{Witten:1992xu}
E.~Witten, \emph{{Two-dimensional gauge theories revisited}}, \href{https://doi.org/10.1016/0393-0440(92)90034-X}{\emph{J. Geom. Phys.} {\bfseries 9} (1992) 303} [\href{https://arxiv.org/abs/hep-th/9204083}{{\ttfamily hep-th/9204083}}].

\bibitem{Griguolo:2024ecw}
L.~Griguolo, R.~Panerai, J.~Papalini, D.~Seminara and I.~Yaakov, \emph{{Localization and resummation of unstable instantons in 2d Yang-Mills}}, \href{https://doi.org/10.1007/JHEP06(2024)188}{\emph{JHEP} {\bfseries 06} (2024) 188} [\href{https://arxiv.org/abs/2403.00053}{{\ttfamily 2403.00053}}].

\bibitem{Closset:2015ohf}
C.~Closset, W.~Gu, B.~Jia and E.~Sharpe, \emph{{Localization of twisted $ \mathcal{N}=\left(0,\;2\right) $ gauged linear sigma models in two dimensions}}, \href{https://doi.org/10.1007/JHEP03(2016)070}{\emph{JHEP} {\bfseries 03} (2016) 070} [\href{https://arxiv.org/abs/1512.08058}{{\ttfamily 1512.08058}}].

\bibitem{Hori:2003ic}
K.~Hori, S.~Katz, A.~Klemm, R.~Pandharipande, R.~Thomas, C.~Vafa et~al., \emph{{Mirror symmetry}}, vol.~1 of \emph{Clay mathematics monographs}, AMS, Providence, USA (2003).

\bibitem{Aspinwall:1991ce}
P.S.~Aspinwall and D.R.~Morrison, \emph{{Topological field theory and rational curves}}, \href{https://doi.org/10.1007/BF02096768}{\emph{Commun. Math. Phys.} {\bfseries 151} (1993) 245} [\href{https://arxiv.org/abs/hep-th/9110048}{{\ttfamily hep-th/9110048}}].

\bibitem{Nieri:2015dts}
F.~Nieri, \emph{{An elliptic Virasoro symmetry in 6d}}, \href{https://doi.org/10.1007/s11005-017-0986-3}{\emph{Lett. Math. Phys.} {\bfseries 107} (2017) 2147} [\href{https://arxiv.org/abs/1511.00574}{{\ttfamily 1511.00574}}].

\bibitem{jeffrey2017}
J.A.~Mracek and L.C.~Jeffrey, \emph{Hyperfunctions, the duistermaat-heckman theorem, and loop groups}, {\emph{Geometry and Physics, Volume I} (2017) } [\href{https://arxiv.org/abs/1706.07388}{{\ttfamily 1706.07388}}].

\bibitem{Benini:2016qnm}
F.~Benini and B.~Le~Floch, \emph{{Supersymmetric localization in two dimensions}}, \href{https://doi.org/10.1088/1751-8121/aa77bb}{\emph{J. Phys. A} {\bfseries 50} (2017) 443003} [\href{https://arxiv.org/abs/1608.02955}{{\ttfamily 1608.02955}}].

\bibitem{Witten:1988xj}
E.~Witten, \emph{{Topological Sigma Models}}, \href{https://doi.org/10.1007/BF01466725}{\emph{Commun. Math. Phys.} {\bfseries 118} (1988) 411}.

\bibitem{Closset:2014pda}
C.~Closset and S.~Cremonesi, \emph{{Comments on $ \mathcal{N} $ = (2, 2) supersymmetry on two-manifolds}}, \href{https://doi.org/10.1007/JHEP07(2014)075}{\emph{JHEP} {\bfseries 07} (2014) 075} [\href{https://arxiv.org/abs/1404.2636}{{\ttfamily 1404.2636}}].

\bibitem{Witten:1993xi}
E.~Witten, \emph{{The Verlinde algebra and the cohomology of the Grassmannian}},  \href{https://arxiv.org/abs/hep-th/9312104}{{\ttfamily hep-th/9312104}}.

\bibitem{Atiyah:1982fa}
M.F.~Atiyah and R.~Bott, \emph{{The Yang-Mills equations over Riemann surfaces}}, {\emph{Phil. Trans. Roy. Soc. Lond. A} {\bfseries 308} (1982) 523}.

\bibitem{GODDARD19771}
P.~Goddard, J.~Nuyts and D.~Olive, \emph{Gauge theories and magnetic charge}, \href{https://doi.org/https://doi.org/10.1016/0550-3213(77)90221-8}{\emph{Nuclear Physics B} {\bfseries 125} (1977) 1}.

\bibitem{Deligne:1999qp}
P.~Deligne, P.~Etingof, D.S.~Freed, L.C.~Jeffrey, D.~Kazhdan, J.W.~Morgan et~al., eds., \emph{{Quantum fields and strings: A course for mathematicians. Vol. 1, 2}} (1999).

\bibitem{Witten:1988hf}
E.~Witten, \emph{{Quantum Field Theory and the Jones Polynomial}}, \href{https://doi.org/10.1007/BF01217730}{\emph{Commun. Math. Phys.} {\bfseries 121} (1989) 351}.

\bibitem{Atiyah:1975jf}
M.F.~Atiyah, V.K.~Patodi and I.M.~Singer, \emph{{Spectral asymmetry and Riemannian Geometry 1}}, \href{https://doi.org/10.1017/S0305004100049410}{\emph{Math. Proc. Cambridge Phil. Soc.} {\bfseries 77} (1975) 43}.

\bibitem{kanwal2012generalized}
R.~Kanwal, \emph{Generalized Functions: Theory and Applications}, Mathematics and Statistics, Birkh{\"a}user Boston (2012), \href{https://doi.org/https://doi.org/10.1007/978-0-8176-8174-6}{https://doi.org/10.1007/978-0-8176-8174-6}.

\bibitem{duistermaat2010distributions}
J.~Duistermaat and J.~Kolk, \emph{Distributions: Theory and Applications}, Cornerstones, Birkh{\"a}user Boston (2010).

\bibitem{Jones1996}
D.S.~Jones, \emph{Hadamard's finite part}, \href{https://doi.org/https://doi.org/10.1002/(SICI)1099-1476(19960910)19:13<1017::AID-MMA723>3.0.CO;2-2}{\emph{Mathematical Methods in the Applied Sciences} {\bfseries 19} (1996) 1017}.

\bibitem{gel2016generalized}
I.~Gel'fand and G.~Shilov, \emph{Generalized Functions, Volume 1}, AMS Chelsea Publishing, American Mathematical Society (2016).

\bibitem{Graf2010}
U.~Graf, \emph{Introduction to Hyperfunctions and Their Integral Transforms: An Applied and Computational Approach}, Birkhauser Basel (2010), \href{https://doi.org/10.1007/978-3-0346-0408-6}{10.1007/978-3-0346-0408-6}.

\bibitem{Imai1992}
I.~Imai, \emph{Applied Hyperfunction Theory}, vol.~8 of \emph{Mathematics and its Applications}, Springer Dordrecht (1992), \href{https://doi.org/10.1007/978-94-011-2548-2}{10.1007/978-94-011-2548-2}.

\bibitem{JeffreyKirwan}
L.C.~Jeffrey and F.C.~Kirwan, \emph{Localization for nonabelian group actions}, \href{https://doi.org/https://doi.org/10.1016/0040-9383(94)00028-J}{\emph{Topology} {\bfseries 34} (1995) 291}.

\bibitem{Cordova:2014oxa}
C.~Cordova and S.-H.~Shao, \emph{{An Index Formula for Supersymmetric Quantum Mechanics}}, \href{https://doi.org/10.5427/jsing.2016.15b}{\emph{J. Singul.} {\bfseries 15} (2016) 14} [\href{https://arxiv.org/abs/1406.7853}{{\ttfamily 1406.7853}}].

\bibitem{Closset:2017bse}
C.~Closset, H.~Kim and B.~Willett, \emph{{$ \mathcal{N} $ = 1 supersymmetric indices and the four-dimensional A-model}}, \href{https://doi.org/10.1007/JHEP08(2017)090}{\emph{JHEP} {\bfseries 08} (2017) 090} [\href{https://arxiv.org/abs/1707.05774}{{\ttfamily 1707.05774}}].

\bibitem{Closset:2017zgf}
C.~Closset, H.~Kim and B.~Willett, \emph{{Supersymmetric partition functions and the three-dimensional A-twist}}, \href{https://doi.org/10.1007/JHEP03(2017)074}{\emph{JHEP} {\bfseries 03} (2017) 074} [\href{https://arxiv.org/abs/1701.03171}{{\ttfamily 1701.03171}}].

\bibitem{Closset:2019hyt}
C.~Closset and H.~Kim, \emph{{Three-dimensional $\mathcal{N}=2$ supersymmetric gauge theories and partition functions on Seifert manifolds: A review}}, \href{https://doi.org/10.1142/S0217751X19300114}{\emph{Int. J. Mod. Phys. A} {\bfseries 34} (2019) 1930011} [\href{https://arxiv.org/abs/1908.08875}{{\ttfamily 1908.08875}}].

\bibitem{Hosseini:2018uzp}
S.M.~Hosseini, I.~Yaakov and A.~Zaffaroni, \emph{{Topologically twisted indices in five dimensions and holography}}, \href{https://doi.org/10.1007/JHEP11(2018)119}{\emph{JHEP} {\bfseries 11} (2018) 119} [\href{https://arxiv.org/abs/1808.06626}{{\ttfamily 1808.06626}}].

\bibitem{Hori:2014tda}
K.~Hori, H.~Kim and P.~Yi, \emph{{Witten Index and Wall Crossing}}, \href{https://doi.org/10.1007/JHEP01(2015)124}{\emph{JHEP} {\bfseries 01} (2015) 124} [\href{https://arxiv.org/abs/1407.2567}{{\ttfamily 1407.2567}}].

\bibitem{Bershtein:2015xfa}
M.~Bershtein, G.~Bonelli, M.~Ronzani and A.~Tanzini, \emph{{Exact results for $ \mathcal{N} $ = 2 supersymmetric gauge theories on compact toric manifolds and equivariant Donaldson invariants}}, \href{https://doi.org/10.1007/JHEP07(2016)023}{\emph{JHEP} {\bfseries 07} (2016) 023} [\href{https://arxiv.org/abs/1509.00267}{{\ttfamily 1509.00267}}].

\end{thebibliography}\endgroup



\providecommand{\href}[2]{#2}\begingroup\raggedright\endgroup



\end{document}